\documentclass[10pt, journal]{IEEEtran}

\usepackage{titlesec}
\usepackage{amsmath}
\usepackage{booktabs}
\usepackage{tabularx}
\usepackage{multirow}
\usepackage{adjustbox} 
\usepackage{caption}
\usepackage{enumitem}
\usepackage{hyperref}
\usepackage{url}
\usepackage{xcolor}
\usepackage{colortbl}
\usepackage{color}
\usepackage{footmisc}
\usepackage[ruled, vlined, linesnumbered]{algorithm2e}
\usepackage{balance}
\usepackage{subcaption}
\usepackage{caption}
\usepackage{acronym}
\usepackage{tcolorbox}
\usepackage{orcidlink}
\usepackage[newfloat,frozencache,cachedir=.]{minted}
\newenvironment{code}{\captionsetup{type=listing}}{}
\SetupFloatingEnvironment{listing}{name=\textbf{Code}}
\tcbuselibrary{skins} 

\definecolor{plotGreen}{HTML}{55a868} 
\definecolor{plotRed}{HTML}{c44e52}  
\definecolor{plotBlue}{HTML}{4c72b0}  

\newcommand{\bz}{\texttt{BLAZE}\xspace}

\newcommand{\bbox}{\texttt{BeetleBox}\xspace}

\def\scf{source code files\xspace}
\def\rqf{How effective is \bz compared to state-of-the-art cross-project bug localization tools?\xspace}

\def\rqs{How effective is \bz compared to state-of-the-art embedding-based bug localization tools?\xspace}

\begin{document}



\title{\bz: Cross-Language and Cross-Project Bug Localization via Dynamic Chunking and Hard Example Learning}
\author{
    \IEEEauthorblockN{Partha Chakraborty\orcidlink{0000-0001-5965-615X}, \IEEEmembership{Student Member, IEEE}, Mahmoud Alfadel\orcidlink{0000-0002-2621-6104}, \IEEEmembership{Member, IEEE},  and Meiyappan Nagappan\orcidlink{0000-0003-4533-4728}~\thanks{Partha Chakraborty and Meiyappan Nagappan is with David R. Cheriton School
of Computer Science, University of Waterloo,  Waterloo, Canada, N2L 3G1. \\E-mail:\{p9chakra, mei.nagappan\}@uwaterloo.ca}
~\thanks{
Mahmoud Alfadel is with the Department of
of Computer Science, University of Calgary,
Calgary, Canada, T2N 1N4. \\E-mail:\ mahmoud.alfadel@ucalgary.ca}
}
}

\IEEEtitleabstractindextext{
\begin{abstract}
Software bugs require developers to expend significant effort to identify and resolve them, often consuming about one-third of their time. 
Bug localization, the process of pinpointing the exact source code files that need modification, is crucial in reducing this effort. Existing bug localization tools, typically reliant on deep learning techniques, face limitations in both cross-project applicability and multi-language environments.

Recent advancements with Large Language Models (LLMs) offer detailed representations for bug localization that may help to overcome such limitations.
However, these models are known to encounter challenges with 1) limited context windows and 2) mapping accuracy.
To address these challenges, we propose \bz, an approach that employs \textit{ dynamic chunking} and \textit{hard example learning}.
First, \bz dynamically segments source code to minimize continuity loss.
Then, \bz fine-tunes a GPT-based model using complex bug reports in order to enhance cross-project and cross-language bug localization.
To support the capability of \bz, we create the \bbox dataset, which comprises 23,782 bugs from 29 large and thriving open-source projects across five programming languages (Java, C$++$, Python, Go, and JavaScript).
Our evaluation of \bz on three benchmark datasets---\bbox, SWE-Bench, and Ye et al.---demonstrates substantial improvements compared to six \textit{state-of-the-art} baselines. 
Specifically, \bz achieves up to an increase of 120\% in Top 1 accuracy, 144\%  in Mean Average Precision (MAP), and 100\% in Mean Reciprocal Rank (MRR).
Furthermore, an extensive ablation study confirms the contributions of our pipeline components to the overall performance enhancement.
 
\end{abstract}
}
\IEEEpubid{\copyright~2025 IEEE. Author pre-print copy. The final publication is available online at: \url{https://doi.org/10.1109/TSE.2025.3579574}}
\maketitle
\IEEEdisplaynontitleabstractindextext

\section{Introduction}
\label{sec:introduction}

Software bugs are deviations from intended functionality in software applications, leading to undesired outcomes~\cite{Koomen1999}. Resolving these bugs requires substantial developer effort, mainly to identify the exact source code file that needs correction. Prior research indicates that debugging consumes about one-third of developer work time~\cite{LaToza2010,Tassey2002}.
The challenge is exacerbated by the often overwhelming number of bug reports compared to the available developer resources, which can negatively affect customer satisfaction~\cite{ElDeeb2023, Zhou2012}.

Bug localization involves pinpointing the specific source code files that need modification based on the details provided in bug reports~\cite{Wang2014}. Effective bug localization tools minimize the time required to identify the root causes of defects by automating the tracing of errors to their sources~\cite{Fischer2003}.

Previous works (e.g., ~\cite{Li2019, Xiao2019, Liang2019, Zhang2020, Zhu2022}) have introduced bug localization tools that measure the similarity between source code files and bug reports. 
These tools, which often employ deep learning techniques like Word2Vec or GloVe embeddings, are effective but heavily reliant on labeled data and generally require retraining for each new project, limiting their cross-project applicability~\cite{Kochhar2014, Sainani2020}.
Although advancements have been made with tools that use transfer learning to function across different projects (e.g., ~\cite{Huo2021, Zhu2020}), they still require labeled data from new projects for retraining. 
Moreover, these tools have predominantly been tested within single-language environments. This raises concerns about their effectiveness in cross-project bug localization involving multiple programming languages.

Recent studies have highlighted that Large Language Models (LLMs) provide detailed and context-rich representations for code retrieval tasks and bug diagnosis~\cite{feng2020, guo2021}. 
This improved representation can enhance the ability of bug 
localization tools to accurately link bug reports with the corresponding \scf by integrating context-sensitive semantic understanding~\cite{Ciborowska2022, Du2023}.
However, such methods still encounter two major challenges.
Firstly (\textbf{Challenge 1: Limited Context Window)}, the typical context window of LLMs, which ranges from 512 to 2,048 tokens, is insufficient when compared to the average token length of 2,536 for a single source code file in common datasets~\cite{Ye2014}.

\begin{figure}[tb!]
  \centering
  \begin{tcolorbox}[colback=white, boxrule=0.8pt, sharp corners, boxsep=0pt, left=1pt, right=1pt, top=1pt, bottom=1pt]
    \begin{subfigure}[t]{0.47\columnwidth}
      \centering
      \includegraphics[width=\linewidth]{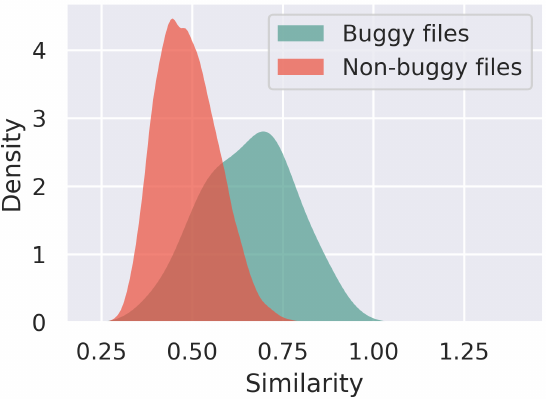}
      \caption{Overlapped distribution before fine-tuning.}
      \label{fig:kde_0}
    \end{subfigure}\hfill 
    \begin{subfigure}[t]{0.47\columnwidth}
      \centering
      \includegraphics[width=\linewidth]{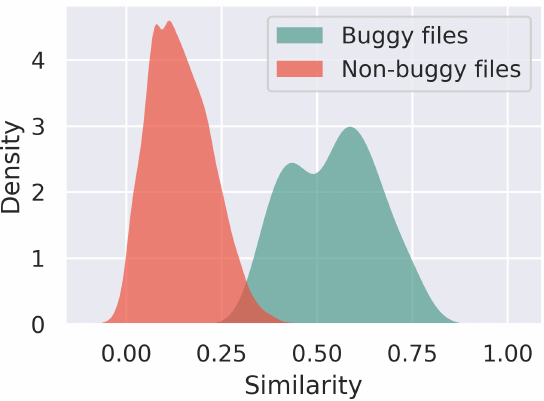}
      \caption{Minimally overlapped distribution after fine-tuning (\bz).}
      \label{fig:kde_15}
    \end{subfigure}
  \end{tcolorbox}
  \label{fig:kdez-plot}
  \caption{Kernel Density Estimate plot showing the distribution of similarities between buggy and non-buggy source code files and bug reports.}
\end{figure}

\begin{figure}[tb!]
  \centering
  \begin{tcolorbox}[colback=white, boxrule=0.8pt, sharp corners, boxsep=2pt, left=2pt, right=2pt, top=2pt, bottom=2pt]
    \includegraphics[width=\columnwidth]{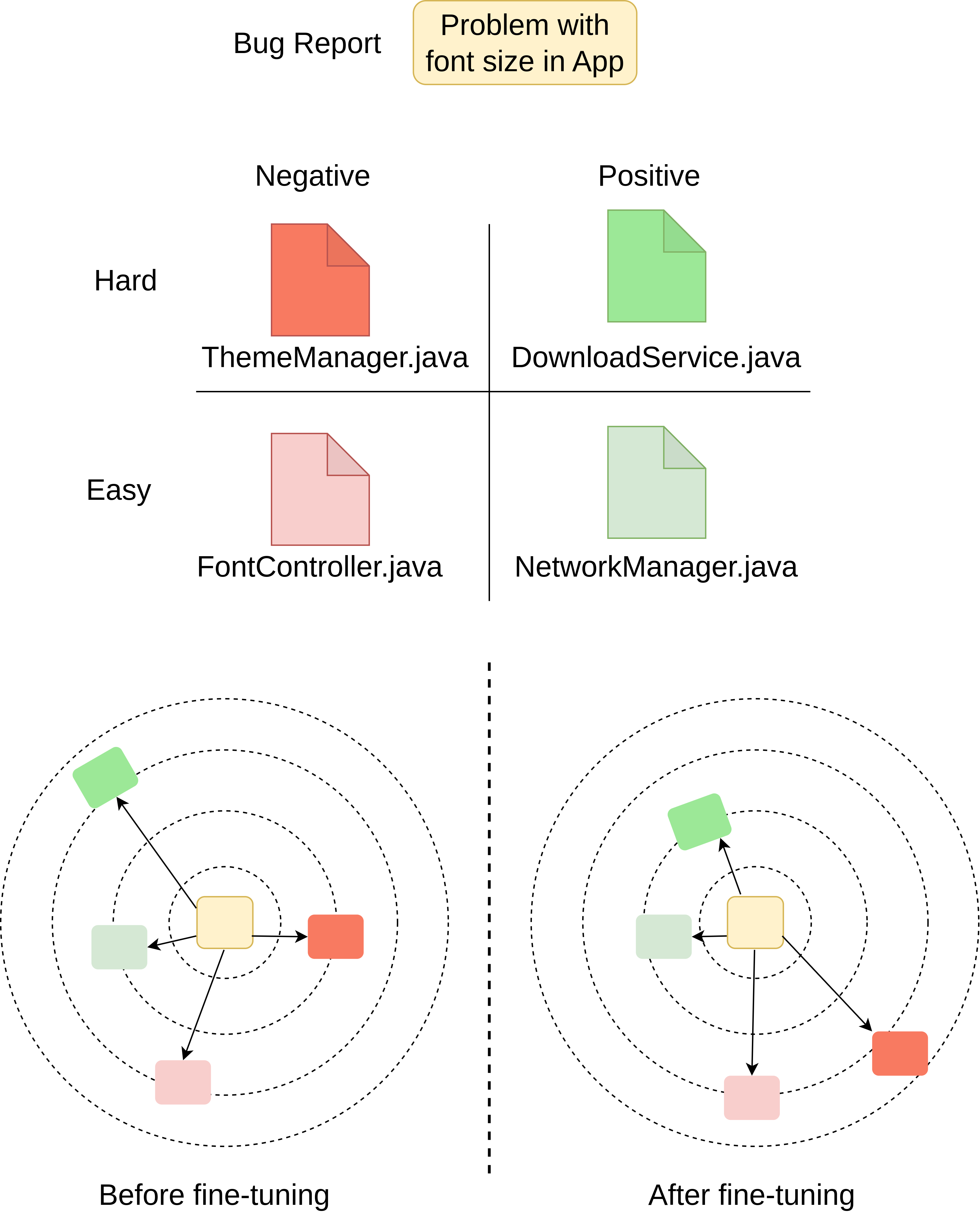}
  \end{tcolorbox}
  \caption{Example of mining hard examples.}
  \label{fig:hard_mining}
\end{figure}

Secondly (\textbf{Challenge 2: Mapping Accuracy}), such methods struggle to accurately map bug reports to the relevant buggy source code file, as illustrated in Figure~\ref{fig:kde_0}, where the ideal non-overlapping distribution between buggy and non-buggy source code is not achieved.
For this challenge, we highlight the need to prioritize the importance of source code files within projects. 
Consider the scenario depicted in Figure~\ref{fig:hard_mining}, where a font issue involves four files. 
The model erroneously associates the bug with `ThemeManager.java,' an irrelevant file, and misses `FontController.java,' which is crucial to the issue. 
Focused learning from such challenging examples can improve the model, aligning the representations of buggy source code files more closely with the bug report.
\newpage

To overcome the above two challenges, we propose \bz, an approach that employs:
\begin{itemize}
    \item[-] \textbf{Dynamic Chunking}: Using dynamic programming, \bz segments source code at the boundaries of source code component declarations (e.g., class, interface, and method) to minimize continuity loss and adhere to context length constraints. 
    We design these segments to be placed at component edges to prevent overlapping, resolving issues of continuity loss and excessive resource consumption caused by repetitive tokens.

    \vspace{2mm}
    \item[-] \textbf{Hard Example Learning:} \bz employs a GPT-based bug localization model tailored for identifying bugs across multiple languages and projects. 
    We fine-tune it by prioritizing complex bugs (hard examples). \bz identifies such cases while tuning out less relevant outliers, which enhances the ability of \bz to generalize across different projects and languages and identifies critical misclassifications~\cite{qiao2023, Xuan2020}. As demonstrated in Figure~\ref{fig:kde_15}, we find that the fine-tuning method could minimize the similarity overlap between buggy and non-buggy code files.
\end{itemize}

\bz localizes bugs at the file level, narrowing the search space for faster bug localization and seamless integration with version control systems~\cite{Ray2017, Zeller2002}. We chose file-level localization for \bz to avoid the complexity (e.g., tracking changes in finer granularity~\cite{liwerski2005,Canfora2005}, hierarchical dependency of changes~\cite{Hassan2004}) and overhead associated with method or commit-level approaches. This balance enhances both accuracy and usability~\cite{Rahman2013, Herzig2013}. Additionally, higher granularity bug localization often leads to more false positives, which negatively impacts developers' experience~\cite{Zimmermann2007}.

To evaluate the capability of \bz across languages, we compile \textsc{\bbox},  the most extensive cross-language and cross-project bug localization dataset to date, to the best of our knowledge. 
\bbox comprises 23,782 bugs that we mine from 29 real-world and thriving projects across five widely used programming languages: C++, Java, JavaScript, Python, and Go.

We evaluate the performance of \bz using three bug localization benchmarks, namely SWE-Bench~\cite{jimenez2024swebench}, Ye et al.~\cite{Ye2014}, and \textsc{\bbox}, which together encompass 30,625 bugs from 36 real-world software projects.  
\bz achieves up to 141\% better Top 1 accuracy and 107.69\% better MAP than  three traditional cross-project bug localization tools~\cite{Huo2021, Zhou2012, Zhu2020}. 
Compared to modern language model-based tools~\cite{Ciborowska2022, Du2023, Neelakantan2022}, \bz achieves improvements of up to 63\% in Top 1 and 69\% in MAP. 
Moreover, we conduct an extensive ablation study, which confirms that our newly designed pipeline and fine-tuning techniques enhance performance. 
The results show that different components of our pipeline substantially contribute to improving bug localization performance to different degrees.

\vspace{2mm}
\noindent
\textbf{Contributions.} In summary, this paper makes the following key contributions:
\begin{itemize}[leftmargin=23pt]
    \item We introduce a novel bug localization technique (i.e., \textsc{\bz}). In this technique,
    \begin{itemize}
         \item We propose a dynamic chunking technique to overcome the limitation of transformer-based approaches (limited context window).
        \item We fine-tune \bz by learning from hard examples. \bz focuses on complex and frequently misclassified cases to enhance model performance in cross-language, cross-project settings.
    \end{itemize}
    \item We create \bbox, the largest to-date cross-language and cross-project bug localization dataset.
    \item We conduct an ablation study to understand the importance of each component of the \bz pipeline.
\end{itemize}

\noindent
\textbf{Fostering open science.} To foster future advances in bug localization, we release our dataset and we make the scripts used for our experiments publicly available~\cite{rep_package}.

\section{Methodology}
\label{sec:methodology}
\renewcommand\thesubsection{Phase \Alph{subsection}}
\titlelabel{\thetitle . \xspace}

In this section, we describe our approach to designing \bz, which consists of three phases: \textbf{Phase A} introduces a fine-tuned language model to associate textual (bug report) and code elements (\scf). 
\textbf{Phase B} implements a dual-indexing system for source code files using both text and vector data from the fine-tuned model to facilitate a more effective retrieval of relevant files. 
Finally, in \textbf{Phase C}, we leverage the fine-tuned model from Phase A  and the resulting indexed databases from Phase B  together to facilitate the full process of matching incoming bug reports with the corresponding source code files. Below, we describe each phase in detail.
\subsection{Fine-tuning of Language Model}
\label{subsec: train_methodology}
 
 \begin{figure}[tb!]
    \centering   
        \includegraphics[scale=0.041]{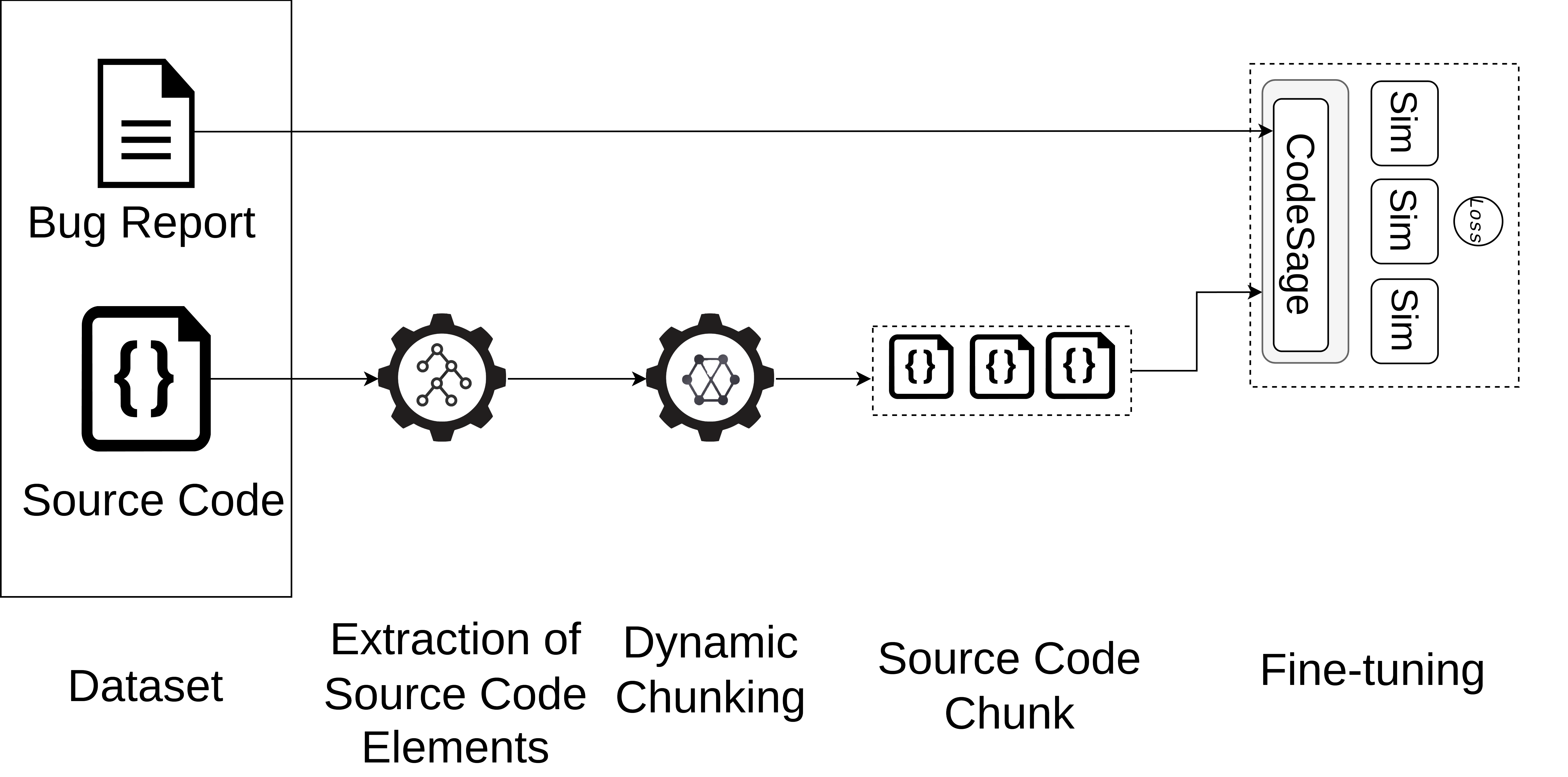}
    \caption{Fine-tuning of language model (Phase A).}
    \label{fig:train_pipeline}
\end{figure}

\noindent Figure~\ref{fig:train_pipeline} presents the steps we follow to fine-tune the model used in \bz. 
Each step of this process is detailed below.
 
\noindent
\textbf{Extraction of Source Code Elements.}
Our process begins with obtaining bug reports, the specific repository version where the bug exists, and all the \scf from that version provided in the targeted dataset (more details on how we collect and process the dataset in this study are described in Section~\ref{subsec:dataset_collection}).
In this initial step, we focus on extracting details such as the names and roles of components (e.g., methods, classes, and interfaces) from the collected source code files. 
Drawing on insights from prior research~\cite{Liu2021}, which indicates that separate indexing of source code files enhances lexical retriever's performance in code search tasks, we index source code files separately in the next phase of our pipeline (Phase B). 
%

Moreover, we extract positional information, such as line numbers, which is essential for subsequent analysis stages. 
To identify all components within the source code, we use tree-sitter,\footnote{\url{https://github.com/tree-sitter/tree-sitter}} a tool that structures code into a graph format. From this tree-sitter graph, we extract all components along with their types, positions (line numbers), and names.

\SetKwProg{Fn}{Function}{}{}
\SetKwInput{Input}{\textbf{Input~}}
\SetKwInOut{Output}{\textbf{Output~}}

\begin{algorithm}

    \caption{Dynamic Chunking.}\label{code:dynamic_chunk}
    \DontPrintSemicolon
    \Input{\;
        \hspace{2pt} cost\_map: Map containing costs for each code component type.\;
        \hspace{2pt} total\_lines: Total number of lines.\;
        \hspace{2pt} window\_size: Maximum chunk size.\;
    }
    \Output{\;
        \hspace{2pt} dp: Array containing minimum costs and corresponding breakpoints\;
    }
    \Fn{dynamic\_chunk(cost\_map, total\_lines, window\_size)}
    {
        line\_map $\leftarrow$ get line number to component type map\;
        split\_cost\_map $\leftarrow$ get line number to cost map\;
        \For{$i \gets$ 1 \KwTo total\_lines}
        {
            minimum cost $\leftarrow$ $INFINITY$\;
            breakpoint $\leftarrow$ $-1$\;
            \For{$j \gets$ MAX(i - window size, 0) \KwTo $i$}
            {
                cost = dp[$j$][0] + split\_cost\_map[$i$] or DEFAULT\_COST\;
                \If{cost $<$ minimum\_cost}
                {
                    minimum\_cost = cost\;
                    breakpoint = $j$\;
                }
            }
            dp[$i$] = (minimum\_cost, breakpoint)
        }
        return dp\;
    }

\end{algorithm}

\noindent
\textbf{Dynamic chunking of source code file.}
In this step, we address the challenge of segmenting lengthy source code files into smaller segments that are suited for context-aware embedding models~\cite{devlin2019, Liu2019, Beltagy2020}.
We employ a dynamic programming approach for chunking. This method segments files at natural boundaries (e.g.,  classes, interfaces, and methods), ensuring continuity and preventing overlaps---issues commonly associated with the traditional static chunking method and the sliding window technique.
%
Algorithm~\ref{code:dynamic_chunk} minimizes continuity loss in code segmentation by using a cost-based dynamic chunking approach. Continuity loss, caused by splitting semantic structures (e.g., methods or classes), is mitigated by assigning lower costs to splits at natural boundaries and higher costs to arbitrary splits. The algorithm uses a cost hierarchy: larger components (e.g., classes) have the lowest costs, smaller ones (e.g., methods) moderate costs, and splits within components have the highest costs (referred as `DEFAULT\_COST' in the algorithm). TreeSitter identifies these boundaries, and a cost dictionary assigns split costs accordingly.

A dynamic programming (DP) array tracks cumulative segmentation costs and optimal breakpoints. The algorithm iteratively evaluates potential split points, selecting those minimizing combined costs of the current segment and previous computations. This cost tracking ensures the minimization of continuity loss. While we are tracking the continuity loss, we also track the current window size, which ensures maximization of resource usage. The final output lists minimal costs and breakpoints, enabling file segmentation aligned with natural code boundaries to preserve semantic integrity. An example of dynamic chunking is available in our online appendix~\cite{rep_package}

\noindent
\textbf{Learning from hard examples.}
\label{susubsec:late interaction}
We fine-tune a Large Language Model (LLM), called CodeSage~\cite{zhang2024codesage}, a GPT-based multi-modal embedding model. CodeSage is adopted to align bug reports and source code files within the same embedding space, leveraging its effectiveness in zero-shot code search tasks~\cite{zhang2024codesage}. However, its effectiveness in bug localization remains limited, as shown in Figure~\ref{fig:kde_0}.

A key distinction between CodeSage and \bz lies in their hard example mining strategies. CodeSage mines examples from docstrings and code, whereas \bz identifies hard examples directly in the embedding space, enabling the detection of challenging cases across diverse programming languages and projects. To enhance CodeSage for bug localization, we employ contrastive learning~\cite{Chen2020}, which minimizes the embedding distance between bug reports and their corresponding buggy files (positive pairs) while maximizing separation from non-buggy files (negative pairs).

To identify hard examples, we first compute a cosine similarity matrix for all possible pairs of bug reports and source code files within a batch. From this matrix, we calculate the median similarity score, which serves as an unbiased threshold, especially when the data distribution is skewed or contains outliers. Next, we identify pairs as hard examples based on the following criteria: (1) positive pairs are considered hard if their similarity score is lower than the median, and (2) negative pairs are considered hard if their similarity score is higher than the median. This approach is grounded in the intuition that these examples deviate from our expectations regarding the similarity between a source code file and a bug report. As such, these pairs require greater attention and emphasis to correct their misalignment with the model’s learned patterns. Since the process identifies samples that are intuitively different from the others within a batch, it is inherently dynamic. As a result, a sample considered “hard” in one batch may not be classified as hard in another batch. An illustration of hard example is presented in Table~\ref{table:hard_example}. As presented in Table~\ref{table:hard_example}, the file listed in the ``Hard Example File Path'' column is not the actual cause of the bug, but it includes terms like “task” that make it seem similar to the issue report. This misleading similarity initially leads \bz to consider it a likely match. However, by treating it as a hard example, \bz learns to recognize false associations during training. Although the list of hard examples is dynamic, the set identified during training is included in our replication package~\cite{rep_package}.


\begin{table}[!tb]
\centering
\caption{Illustration of hard example.}
\label{table:hard_example}
\begin{tabular}{@{}lp{3cm}p{3cm}@{}}
\toprule
\textbf{Issue Id} & \textbf{Issue Title}                                                     & \textbf{Hard Example File Path}                      \\ \midrule
\href{https://github.com/apache/dolphinscheduler/issues/11003}{11003}             & {[}TaskGroupOption{]} Task group queue is not updated to the final state & \url{airflow/providers/google/cloud/operators/bigquery.py} \\ \bottomrule
\end{tabular}

\vspace{1em}
\noindent\textbf{Hard Example File Content (code):}
\begin{code}
\label{code:code4}
\begin{minted}[breaklines,tabsize=1, highlightlines={9,10,15,16},fontsize=\footnotesize]{py}
class BigQueryConsoleLink(BaseOperatorLink):
"""
Helper class for constructing BigQuery link.
"""

name = 'BigQuery Console'

    def get_link(self, operator, dttm):
        ti = TaskInstance(task=operator, execution_date=dttm)
        job_id = ti.xcom_pull(task_ids=operator.task_id, key='job_id')
        return BIGQUERY_JOB_DETAILS_LINK_FMT.format(job_id
        =job_id) if job_id else ''
    
    ....
    def get_link(self, operator: BaseOperator, dttm: datetime):
            ti = TaskInstance(task=operator, execution_date=dttm)
            job_ids = ti.xcom_pull(task_ids=operator.task_id, key='job_id')
            if not job_ids:
                return None
            if len(job_ids) < self.index:
                return None
            job_id = job_ids[self.index]
            return BIGQUERY_JOB_DETAILS_LINK_FMT.format
            (job_id=job_id)
        
\end{minted}
\end{code} 
\end{table}

The \bz training is guided by a modified version of Normalized Temperature-scaled Cross-Entropy (NT-Xent) loss~\cite{Chen2020}, defined as:

\begin{align}
\label{eq:hardntxentloos}
\begin{gathered}
\mathcal{L}_{i,j} = -\log\frac{M_{i,j} * \exp^{sim(z_i, z_j)/\tau}}{\sum_{k=1}^{N} 1_{\lbrack k\neq i \rbrack}\exp^{sim(z_i, z_k)/\tau} * M_{i,k}}
    \\
    M_{m,n} = \begin{cases}
        \alpha, & \text{\parbox[t]{.3\textwidth}{if $z_n$ negative and  $sim(z_m, z_n)$ $>$ $median(sim(z_u, z_v))$}}\\
        \beta, & \text{\parbox[t]{.3\textwidth}{if $z_n$ positive and  $sim(z_m, z_n)$ $<$ $median(sim(z_u, z_v))$}}\\
        1, & \text{otherwise}
    \end{cases}\\
    \alpha > 1;\beta > 1; 1\leq u,v\leq N
\end{gathered}
\end{align}

\indent where $z_i$ is the embedding of the bug report, $z_j$ is the embedding of the positive (buggy) source code file, and $\tau$ is the temperature parameter that scales the similarity. 
This loss function distinguishes between positive and negative examples by comparing similarity scores. As explained previously, we identify the hard examples, and in the loss function, they are prioritized by $\alpha$ and $\beta$.

In an average project, approximately 700 source code files exist, with only 1-2 directly related to a reported bug, as evidenced in Apache project dataset~\cite{Ye2014}. 
This rarity of relevant files presents a considerable challenge in bug localization. 
Our approach of hard example mining and scaling enhances the gradients from challenging cases and helps maintain a balanced learning environment, preventing the model from being overwhelmed by outliers. 
Moreover, the continuous nature of the scaling multiplier preserves the convergence properties of the NT-Xent loss. By employing median-based scaling, our method could improve the model learning from difficult examples, thereby enhancing its convergence. Henceforth, we refer to the fine-tuned language model as the \bz model.

\begin{figure}[tb!]
    \centering   
        \includegraphics[scale=0.021]{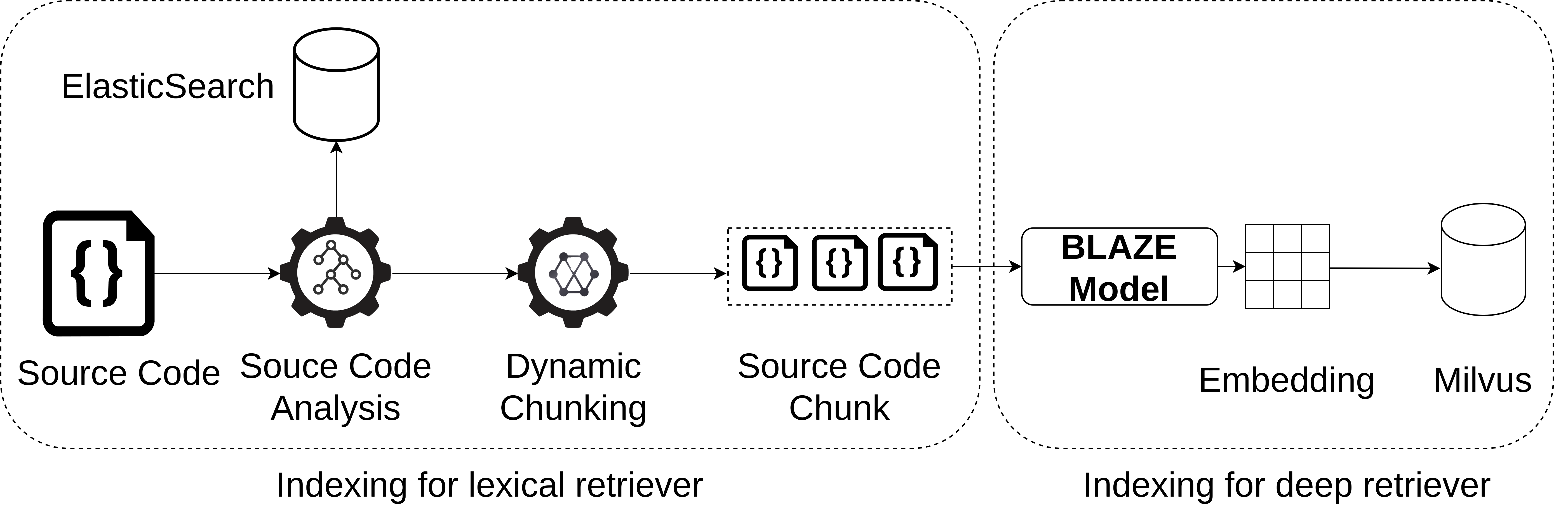}
    \caption{Offline indexing phase of \bz (Phase B).}
    \label{fig:index_pipeline}
\end{figure}

\subsection{Offline Indexing}
\label{subsec:piepline_methodology}
We introduce the offline indexing phase to enhance bug localization by proactively processing and indexing \scf. 
As mentioned earlier, repositories contain (on average) about 700 \scf, and without this phase, developers could face delays due to real-time data processing.

As depicted in Figure~\ref{fig:index_pipeline}, upon receiving a bug report, \bz employs two distinct indexing strategies: the \textit{deep retriever} stores data in the Milvus vector database~\cite{Wang2021}, while the \textit{lexical retriever} uses the no-SQL database ElasticSearch~\cite{elasticsearch}. Having two types of retrievers can help to offset their weaknesses and take advantage of their strengths. Note that this phase builds upon the earlier phase of extracting names and positions of components (see \ref{subsec: train_methodology}).
Below, we describe each retriever in detail.

\noindent\textbf{Indexing for lexical retriever.}
Lexical retrievers are fast and use keyword matching to assess similarity. 
After extracting component names and positions as detailed in~\ref{subsec: train_methodology}), each source code file is indexed in ElasticSearch. 
ElasticSearch leverages the Okapi BM25 algorithm~\cite{bm25} to compute similarity scores, a technique well-documented in previous studies for applications in code retrieval~\cite{Liu2021, Sachdev2018}.
Additionally, we index filenames and file paths to match logs and stack traces within bug reports.

\noindent
\textbf{Indexing for deep retriever.}
Deep retrievers integrate semantics to calculate the similarity between bug reports and \scf.
Starting with the dynamically chunked source code files---as outlined in ~\ref{subsec: train_methodology}---each chunk is transformed into an embedding using the \bz model. 
These embeddings are then indexed in Milvus, an open-source vector database specifically optimized for storing and retrieving embedding vectors in artificial intelligence applications.


\subsection{Retrieval Pipeline}
 \begin{figure}[tb!]
    \centering   \includegraphics[scale=0.028]{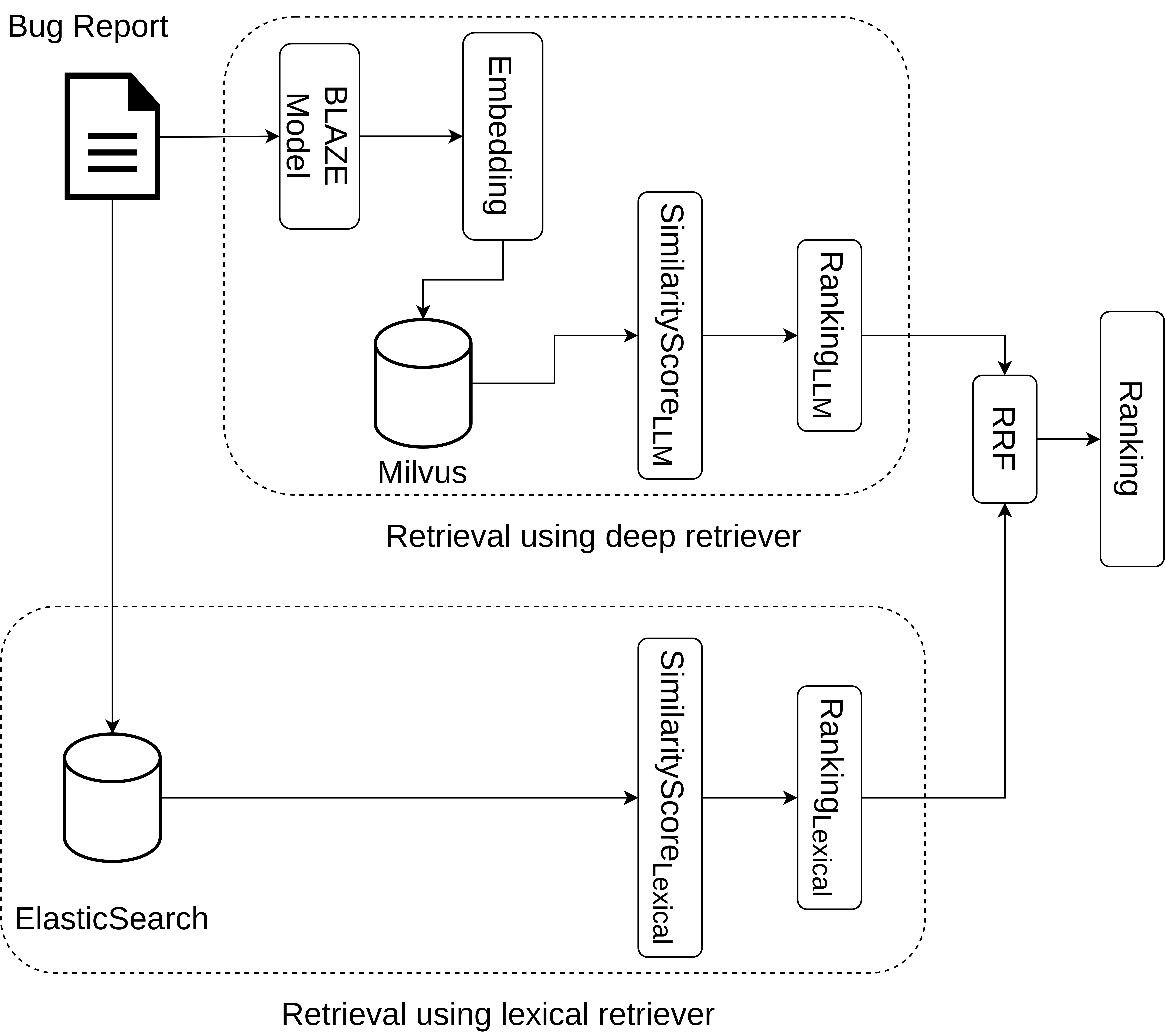}
    \caption{Retrieval phase of \bz (Phase C).}
    \label{fig:retrieval_pipeline}
\end{figure}
Figure~\ref{fig:retrieval_pipeline} illustrates the retrieval phase of \bz.
We leverage a combination of lexical and deep retrievers along with the fine-tuned \bz model.
As shown in the figure, upon the arrival of a bug report, we query the Milvus and ElasticSearch indices to pinpoint the relevant \scf. 
We then integrate the rankings from both indices using a technique known as Reciprocal Rank Fusion (RRF).
Next, we provide a detailed description of these steps.

\noindent
\textbf{Retrieval using deep retriever.}
Deep retrievers have demonstrated superior performance in ranking tasks compared to traditional vocabulary matching approaches~\cite{Abbasiantaeb2021}. 
To use a deep retriever, we first calculate the embedding of the bug report using our fine-tuned \bz model. 
Then, we retrieve the embedding of all the source code files from Milvus, a vector index we built during offline indexing.
Milvus delivers the top K most similar chunks to the bug report embedding. 
We use cosine similarity to assess the similarity between embeddings. Finally, we sort the source code files based on decreasing similarity. 

\noindent
\textbf{Retrieval using lexical retriever.}
In the lexical retrieval process, we query the ElasticSearch index with the bug report, which returns similarity scores for each source code file based on lexical matches. 
We use these scores to rank the source code files, leveraging ElasticSearch text searching capabilities to localize files potentially containing the bug. 
By directly using the bug report as a query, we ensure that the retrieval is tightly focused on the textual content relevant to the bug.

\noindent
\textbf{Reciprocal rank fusion.}
While deep retrievers demonstrate superior performance in capturing semantic similarity, they sometimes struggle with lengthy documents and may face challenges such as domain shifts when applied across different projects~\cite{Luan2021, Wangb2021}. These shifts can lead to outdated understandings of new domains. Conversely, lexical retrievers are immune to domain shifts but can encounter difficulties with vocabulary mismatches due to their lack of a semantic understanding. To address these limitations and harness the strengths of both retrieval types, we employ Reciprocal Rank Fusion (RRF)~\cite{Cormack2009}.
RRF integrates the rankings from both the lexical and deep retrievers by considering the position of each source code file in the rankings rather than merging their numerical scores.
This method, supported by prior research~\cite{Cormack2009, Bendersky2020, Chen2022rrf}, has proven to be a reliable and powerful technique for creating robust ensembles that enhance overall retrieval performance.

Given a bug report $r$, a source code file $s$, and a set of retrievers $M$, where $\pi^m(r,s)$ represents the ranking by retriever $m$ for source code file $s$ and bug report $r$, the RRF score is defined as follows:

\begin{equation}
RRF(r,s, M) = \sum_{m\epsilon M}\frac{1}{k + \pi^m(r, s)}
\end{equation}

Here, $k$ is a hyperparameter.
Based on prior studies~\cite{Cormack2009, Chen2022rrf}, $k=60$ has been found to provide the highest mean average precision (MAP).
We also experimented with $k$ values ranging  $\epsilon [50, 70]$ in the \bbox dataset and confirmed that $k=60$ continues to yield the highest MAP.
\renewcommand\thesubsection{\thesection.\Alph{subsection}}
\titlelabel{\thetitle \quad}
\section{Experimental Design}
In this section, we present the Research Questions (RQs) we tackle in our study, the process of our dataset collection, the subject baselines, and the performance metrics employed to evaluate the performance of \bz.
\label{sec:experimental_evaluation}
\subsection{Research  Questions}
Below, we present our RQs by explaining the motivation behind each one.

\noindent
\textbf{RQ1.} \textit{\rqf\\}
\noindent
A plethora of bug localization tools have been proposed (e.g., ~\cite{Zhu2022, Xiao2018, Huo2016, Zhu2020}), with cross-project tools (e.g.,~\cite{Huo2021, Zhu2020}) being particularly effective. However, these tools still require project-specific additional training. 
Our proposed approach, \bz, is designed to function without such a requirement. This RQ evaluates how well \bz performs compared to state-of-the-art cross-project bug localization tools.
\noindent
\textbf{RQ2.} 
\textit{\rqs\\}
\noindent
Embedding-based techniques, which assess document similarity, are extensively used for bug localization. 
These techniques differ from traditional bug localizers that provide binary predictions by producing embeddings and employing metrics like cosine distance to evaluate if a source code file contains bugs. 
Their ability to contextualize associations between bug reports and source code enhances their effectiveness. This RQ explores how well \bz performs compared to these embedding-based tools.

\subsection{Dataset Curation of \textsc{\bbox}}
\label{subsec:dataset_collection}
We have compiled an extensive multi-language, multi-project bug localization dataset. 
Previous research~\cite{Ye2014, Zhou2012,Moreno2014,Dit2011} has proposed various datasets for bug localization, but these have numerous shortcomings, such as support for only one programming language and a limited range of projects. Additionally, some datasets, such as Bench4BL~\cite{Lee2018}, contain inaccuracies in their ground truth data~\cite{Kim2021}. 
To address these challenges, we introduce \textsc{\bbox}. 
Below, we detail the process we follow for curating \bbox.
\begin{itemize}[leftmargin=2em]

    \item  \textbf{Selection of repositories.} We use the GitHub API search feature to collect repository names. We search for repositories written in the top five programming languages, Java, Python, C$++$, JavaScript, and Go, as ranked in the StackOverflow Survey.\footnote{\url{https://survey.stackoverflow.co/2022/\#technology-most-popular-technologies}}
    We select the top 20 repositories for each language, sorting them in descending order by their star ratings and most recent updates. 
    This step guarantees that the repositories are actively contributed to and popular within the open-source community.
    \item \textbf{Linking pull-requests and issues.} 
    We gather `closed' issues for each repository and their corresponding fixing Pull Request (PR). 
    We identify these PRs by using predefined GitHub keywords.\footnote{\url{https://docs.github.com/en/issues/tracking-your-work-with-issues/linking-a-pull-request-to-an-issue}}
    \item \textbf{Filtration of issues.} An issue may have several PRs, but typically only one is merged. To minimize errors in our ground truth, we filter out issues with the `duplicate' tag and verify using GitHub API that each issue is linked to a single, merged PR. We initially started with 100 repositories. However, in certain projects, we found no issues meeting all the criteria. Additionally, to prevent possible data leakage~\cite{elangovan2021}, where pretrained LLMs might have seen  evaluation data, we applied a two-step repository-level filtering approach,
    \begin{itemize}
        \item \textbf{Repository Identification:} We cross-referenced all candidate repositories with the training data of the CodeSage model~\cite{zhang2024codesage} used in our study (as disclosed by the model provider or inferred through public documentation).
        \item \textbf{Repository Removal:} If a repository was identified as being part of the LLM’s pretraining corpus, we removed all bug reports and associated files from that repository from our dataset.
    \end{itemize}
    While we acknowledge that finer-grained filtering (e.g., removing only specific bug reports seen during training) would allow greater control, this would require detailed access to pretraining data, which is often unavailable and computationally intensive to trace. Therefore, we chose the more conservative repository-level removal strategy.

As a result of this process, we filtered out 63 repositories, leaving a final dataset of 29 repositories across five programming languages. This ensures that none of the testing data overlaps with the LLM’s training set. Furthermore, in certain cases, we identified mentions of the buggy files within the issue body. To ensure the quality of the dataset, we also removed those issues.
\end{itemize}

For each bug report in \bbox, the dataset, we include the bug status, repository name, repository URL, issue ID, a list of files modified during the fix, the title of the bug report, its body, URL link to the merged PR, URL link to the issue, SHA values of before and after the fix, the date and time the bug was reported, and the date and time of the fixing commit. For matching bug reports with source code files, we use the multiple version set matching approach~\cite{Lee2018}, which uses the SHA of the before-fix commit to identify the source code files of that version. As shown in prior research~\cite{Lee2018}, this method reflects developers' bug localization practices and improves tool performance by reducing dataset noise.
Overall, \bbox comprises 23,782 bugs extracted from 29 projects. 
Detailed statistics about \bbox and a list of the projects can be found in the online Appendix~\cite{rep_package}.

Finally, to assess the accuracy of the ground truth in \bbox, we use the method outlined by Kim et al.~\cite{Kim2021}.
Our manual analysis reveals that \bbox has an error rate of 0.06\%, with a 95\% confidence level.


\subsection{Evaluation Benchmarks and Baselines}

\noindent
\textbf{Benchmarks.}
In addition to our \bbox dataset, we employ SWE-Bench~\cite{jimenez2024swebench}, and Ye et al.~\cite{Ye2014}.
Similar to \bbox, Ye et al. dataset is also recognized for its low error rate in its ground truth. 
While \bbox and Ye et al. are solely bug localization datasets, SWE-Bench is a recent dataset that is a benchmarking suite for bug localization as well as automatic patch generation.

\begin{table}[tb!]
\caption{Statistics on the used datasets, showing the distribution of bugs across training and testing, categorized by language for each dataset.}
\centering
\resizebox{0.77\columnwidth}{!}{%
\begin{tabular}{lllr}
\toprule
\textbf{Dataset}            & \textbf{Purpose}       & \textbf{Language} & \textbf{\# Bugs} \\ \midrule
\multirow{10}{*}{\bbox} & \multirow{5}{*}{train} & c$++$               & 3,744             \\ \cline{3-4} 
                            &                        & go                & 727              \\ \cline{3-4} 
                            &                        & java              & 2,553             \\ \cline{3-4} 
                            &                        & javascript        & 1,703             \\ \cline{3-4} 
                            &                        & python            & 3,034             \\ \cline{2-4} 
                            & \multirow{5}{*}{test}  & c$++$               & 4,407             \\ \cline{3-4} 
                            &                        & go                & 384              \\ \cline{3-4} 
                            &                        & java              & 2,058             \\ \cline{3-4} 
                            &                        & javascript        & 2,963             \\ \cline{3-4} 
                            &                        & python            & 2,209             \\ \hline
SWE-Bench~\cite{jimenez2024swebench}                   & test                   & python            & 2,294             \\ \hline
Ye et al.~\cite{Ye2014}                   & test                   & java              & 16,314            \\ \bottomrule
\end{tabular}
}
\label{table:data_stat}
\end{table}
In our methodology, we fine-tune the language model using the training split of \bbox.
For evaluation, we use the test splits of \bbox, SWE-Bench, and the dataset from Ye et al.
Table~\ref{table:data_stat} provides brief statistics of these three datasets.

\noindent
\textbf{Studied baselines.}
We compare \bz with state-of-the-art cross-project bug localization tools, TRANP-CNN~\cite{Huo2021}  and CooBa~\cite{Zhu2020}, which both use few-shot learning for cross-project adaptation. TRANP-CNN, a CNN-based deep learning tool, and CooBa, which uses dual encoders, train on a source project and are fine-tuned on a target project with limited samples. 
We also include BugLocator~\cite{Zhou2012} as a baseline in our evaluation.
Although not designed for cross-project use, BugLocator, like \bz, is an information retrieval-based tool that does not require additional training for deployment in new projects. The characteristics make it a relevant comparison in our evaluation.

In addition, we evaluate the performance of \bz compared to embedding-based bug localizers, namely SemanticCodeBERT~\cite{Du2023} and FBL-BERT~\cite{Ciborowska2022}.
Both are BERT-based models, where SemanticCodeBERT incorporates semantic flow graph information into its training, and FBL-BERT incorporates changeset data. Recent advancements in generative AI have also introduced models like \emph{openai-embedding}~\cite{Neelakantan2022}, which have demonstrated superior performance in tasks involving natural language and code. Therefore, OpenAI embedding is also included in our baseline for comparison. Of the six baseline tools, we use the bug report and source code file as text input for BugLocator, TRANP-CNN, CooBA, and OpenAI. For SemanticCodeBERT, we process the source code to extract the Semantic Flow Graph (SFG) as recommended in the original study~\cite{Du2023}. Finally, for FBL-BERT, we calculate the changeset to use it as input. While the input formats vary slightly, all tools use the same dataset and attributes to produce these different input representations (e.g., SFG).

\subsection{Evaluation Metrics}
We adopt the following metrics that are widely adopted for evaluating bug localization tools~\cite{Huo2021, Huo2016, Zhou2012}: Top N, Mean Reciprocal Rank (MRR), and Mean Average Precision (MAP).

\begin{itemize}[leftmargin=2em]
    \item \textbf{Top N.}
Top N measures the overall ranking performance of the bug localization model. It indicates the percentage of bug reports for which at least one buggy source code file appears among the top N positions in the ranked list generated by the bug localization tool. 
Following previous studies~\cite{Huo2016, Huo2021}, we consider three values of N: 1, 5, and 10.

\item \textbf{Mean Reciprocal Rank (MRR).}
MRR measures the average rank of the relevant file in the retrieved files set. 
It calculates the average reciprocal rank of the relevant source code files across all bug reports. 
The following equation calculates MRR, where $A$ represents the set of bug reports.

\begin{equation}
MRR= \frac{1}{|A|}\sum_{A}\frac{1}{Least\ rank\ of\ the\ relevant\ files}
\end{equation}

For each bug report, our bug localization model ranks the \scf. 
Let us consider two bug reports, $report_1$ and $report_2$, each with six candidate \scf ranked by the model. 
For $report_1$, the ground truth of the retrieved files are $[0, 0, 1, 0, 1, 0]$, and for $report_2$, the ground truth of the retrieved files are $[1, 0, 0, 0, 0, 1]$. In this case, the least rank of relevant files is 3 and 1, respectively, for $report_1$ and $report_2$. Therefore, $MRR = \frac{1}{2}(\frac{1}{3} + \frac{1}{1}) = 0.67$.

\item \textbf{Mean Average Precision (MAP).}
MAP measures the retrieval quality in scenarios where a bug is linked to multiple \scf. 
MAP considers the rank of all relevant files in the retrieved files list, making it a more comprehensive and unbiased metric than MRR. 
To compute MAP, we determine precision@1, precision@2, ..., and precision@k; then, we calculate the average precision at different points.
After calculating the average precision for each bug report, we calculate the mean of the average precision to compute MAP. The equation for calculating MAP is presented below, where $A$ represents the set of bug reports.

\begin{equation}
MAP = \frac{1}{|A|}\sum_{A}{AvgPrecision(Report_i)}
\end{equation}

To demonstrate the computation of MAP, consider the same example of two bug reports as previously discussed. 
The average precision calculated for $report_1$ is 0.37, while it is 0.67 for $report_2$.
Thus, $MAP = \frac{1}{2}(0.36 + 0.67) = 0.52$.
\end{itemize}

\subsection{Configuration Setup}
We fine-tune the \bz model in a distributed setting with eight cores, 128GB of RAM, and fourteen Nvidia V100 32GB GPUs. 
We conduct inference on the same server using an Nvidia V100 32 GB GPU, which allows for fast and efficient predictions.
We use PyTorch v.2.1.2, PyTorch-lightning v2.2.0, and the HuggingFace library v.4.36.2 to fine-tune our model. 
We also use ElasticSearch v8.7.1 and Milvus v2.3.4  for efficient similarity search. 
We fine-tune our language model on the \bbox dataset for 15 epochs, with batches of size 80 and a decaying learning rate starting at 2E-07, as recommended in prior work~\cite{devlin2019, Ciborowska2022}. 
More details about the hyperparameters are available in our online Appendix~\cite{rep_package}.

\section{Results}
\label{sec:results}
In this section, we present our results per each RQ.

\subsection*{RQ1: Comparison to Cross-Project Tools.}
\begin{table}[tb!]
\caption{Comparative performance of \bz and cross-project bug localization tools.}
\label{table:zero_shot}
\centering
\resizebox{\columnwidth}{!}{%
\begin{tabular}{llrrrrr}
\hline
\textbf{Dataset}           & \textbf{Model Name} & \textbf{Top 1} & \textbf{Top 5} & \textbf{Top 10} & \textbf{MAP}  & \textbf{MRR}  \\ \hline
\multirow{4}{*}{\bbox}     & \bz                 & \textbf{0.18}  & \textbf{0.37}  & \textbf{0.45}   & \textbf{0.22} & \textbf{0.27} \\
                           & TRANP-CNN           & 0.08           & 0.13           & 0.19            & 0.09          & 0.13          \\
                           & CooBa               & 0.06           & 0.1            & 0.14            & 0.08          & 0.09          \\
                           & BugLocator          & 0.06           & 0.10           & 0.15            & 0.08          & 0.12          \\ \hline
\multirow{4}{*}{SWE-Bench} & \bz                 & \textbf{0.29}  & \textbf{0.68}  & \textbf{0.79}   & \textbf{0.43} & \textbf{0.45} \\
                           & TRANP-CNN           & 0.13           & 0.28           & 0.42            & 0.26          & 0.34          \\
                           & CooBa               & 0.12           & 0.25           & 0.38            & 0.21          & 0.3           \\
                           & BugLocator          & 0.1            & 0.14           & 0.21            & 0.17          & 0.21          \\ \hline
\multirow{4}{*}{Ye et al.} & \bz                 & \textbf{0.16}  & \textbf{0.35}  & \textbf{0.44}   & \textbf{0.21} & \textbf{0.26} \\
                           & TRANP-CNN           & 0.09           & 0.17           & 0.23            & 0.14          & 0.15          \\
                           & CooBa               & 0.08           & 0.13           & 0.19            & 0.09          & 0.13          \\
                           & BugLocator          & 0.09           & 0.11           & 0.16            & 0.07          & 0.12          \\ \hline
\end{tabular}
}
\end{table}

\noindent\textbf{Approach.} 
We train our model using \bbox on 11K bugs from five languages, covering 13 projects. Our evaluation dataset contains 30,625 bugs from five languages covering 29 real-world software projects.
Following prior studies~\cite{jimenez2024swebench, Fan2024}, \bbox uses a completely different set of projects for training and testing splits.
As previously presented in Table~\ref{table:data_stat},  the SWE-Bench and Ye et al. datasets are used only for testing.

\vspace{1mm}
\noindent\textbf{Result.} 
Table~\ref{table:zero_shot} presents the performance of \bz compared to TRANP-CNN, CooBa, and BugLocator, across the studied datasets. Overall, \bz demonstrates its highest performance in the SWE-Bench dataset. Conversely, its performance is reduced in the Ye et al. dataset, as indicated by both Top 1 and MAP scores. Nonetheless, \textbf{\bz consistently outperforms comparative baselines in this dataset and across all evaluated datasets.} 

In the \textbf{\bbox dataset}, \bz surpasses TRANP-CNN, with improvements ranging between 107\% to 144.44\%. 
Specifically, \bz achieves a 125\% increase in Top 1 accuracy, a 144.44\% increase in MAP, and a 107.69\% in MRR.
When compared to CooBa, \bz shows substantial gains: 200\% in Top 1, 175\% in MAP, and 200\% in MRR.
Against BugLocator, \bz shows improvements of 200\% in Top 1, 175\% in MAP, and 125\% in MRR.

In the \textbf{SWE-Bench dataset}, 
\bz performs better than TRANP-CNN with increases between 32\% and 123\% across various metrics.
The enhancements include a 123\% improvement in Top 1, 65\% in MAP, and 32\% in MRR.
Against CooBa, \bz shows a 141\% increase in Top 1, 104\% in MAP, and 50\% in MRR.
Compared to BugLocator, \bz shows improvements of 190\% in Top 1, 52.9\% in MAP, and 61.9\% in MRR.

In the \textbf{Ye et al. dataset}, \bz shows an improvement over TRANP-CNN with gains of 77\% in Top 1, 50\% in MAP, and 73\% in MRR.
In comparison to CooBa, \bz shows increases of 100\% in Top 1, 133\% in MAP, and 100\% in MRR.
Against BugLocator, the model posts gains of 77.7\% in Top 1, 200\% in MAP, and 25\% in MRR.\par

Additionally, across the studied programming languages, \bz achieves Top 1 scores of 0.18 in C$++$, 0.36 in Go, 0.26 in Java, 0.21 in JavaScript, and 0.3 in Python.
These results show an improvement of 50.9\% to 135.6\% compared to TRANP-CNN, with the highest improvement observed in  JavaScript and the lowest in Python. 
The trend is similar to that of CooBa, with improvements ranging from 82\% to 162\%, with the highest increase observed in C$++$ and the lowest in Python. Against BugLocator, the improvement ranges from 87\% to 160\%, with the highest improvement in JavaScript and the lowest in Python.
\subsection*{RQ2: Comparison to Embedding-Based Tools.}

\setlength{\tabcolsep}{2pt}

\begin{table}[tb!]
\caption{Comparative performance of \bz and embedding-based bug localization tools.}
\centering
\resizebox{\columnwidth}{!}{%
\begin{tabular}{llrrrrr}
\hline
\textbf{Dataset}           & \textbf{Model Name} & \textbf{Top 1} & \textbf{Top 5} & \textbf{Top 10} & \textbf{MAP}  & \textbf{MRR}  \\ \hline
\multirow{4}{*}{\bbox}     & \bz                 & \textbf{0.18}  & \textbf{0.37}  & \textbf{0.45}   & \textbf{0.22} & \textbf{0.27} \\
                           & OpenAI              & 0.13           & 0.26           & 0.32            & 0.1           & 0.19          \\
                           & SemanticCodeBERT    & 0.11           & 0.20           & 0.31            & 0.13          & 0.21          \\
                           & FBL-BERT            & 0.12           & 0.23           & 0.30            & 0.16          & 0.19          \\ \hline
\multirow{4}{*}{SWE-Bench} & \bz                 & \textbf{0.29}  & \textbf{0.68}  & \textbf{0.79}   & \textbf{0.43} & \textbf{0.45} \\
                           & OpenAI              & 0.27           & 0.63           & 0.72            & 0.36          & 0.39          \\
                           & SemanticCodeBERT    & 0.23           & 0.5            & 0.64            & 0.31          & 0.33          \\
                           & FBL-BERT            & 0.20           & 0.57           & 0.51            & 0.31          & 0.30          \\ \hline
\multirow{4}{*}{Ye et al.} & \bz                 & 0.16           & 0.35           & \textbf{0.44}   & \textbf{0.21} & \textbf{0.26} \\
                           & OpenAI              & \textbf{0.17}  & \textbf{0.36}  & 0.43            & 0.19          & 0.25          \\
                           & SemanticCodeBERT    & 0.13           & 0.24           & 0.30            & 0.17          & 0.22          \\
                           & FBL-BERT            & 0.10           & 0.24           & 0.28            & 0.14          & 0.18          \\ \hline
\end{tabular}
im}
\label{table:embedding_comparison}
\end{table}

\noindent\textbf{Approach.}
We examine three embedding-based bug localization tools: SemanticCodeBERT~\cite{Du2023}, FBL-BERT~\cite{Ciborowska2022}, and OpenAI~\cite{Neelakantan2022}.
For SemanticCodeBERT\footnote{\url{https://github.com/IBM/semanticflowgraph}} and FBL-BERT\footnote{\url{https://anonymous.4open.science/r/fbl-bert-700C/README.md}}, we use their publicly available replication package to train the models and produce embeddings.
For OpenAI, we obtain embeddings for both bug reports and source code files using the OpenAI API.\footnote{\url{https://openai.com/index/openai-api/}} 
We employ cosine similarity to rank the source code files based on their similarity scores, facilitating the identification of the buggy files.

\vspace{1mm}
\noindent\textbf{Result.} 
Table~\ref{table:embedding_comparison} illustrates the performance of \bz relative to the studied embedding-based tools.
From the table, \textbf{we observe trends similar to those observed in Table~\ref{table:zero_shot}}.
We also observe that the performance across all models is strongest in the SWE-Bench dataset, which comprises Python projects, and weakest in the Ye et al. dataset, featuring Java projects.

In the \textbf{\bbox dataset.}, \bz demonstrates superior performance to OpenAI, with improvements ranging from 38\% in Top 1 accuracy to 120\% in MAP and 42.11\% in MRR. Similarly, when compared to SemanticCodeBERT, \bz achieves substantial gains, with increases of 63\% in Top 1, 69\% in MAP, and 28\% in MRR. It also shows notable enhancements against FBL-BERT, including a 50\% rise in Top 1 and improvements in MAP and MRR by 37\% and 42\%, respectively.

The performance trend continues in the \textbf{SWE-Bench dataset}, where \bz outperfomrs OpenAI with gains from 7\% in Top 1 to 19\% in MAP and 15\% in MRR.
It also outperforms SemanticCodeBERT with a 26\% increase in Top 1 and even greater enhancements in MAP and MRR. 
Against FBL-BERT, \bz shows substantial improvements of 45\% in Top 1, 38\% in MAP, and 50\% in MRR.

In the \textbf{Ye et al. dataset}, \bz slightly lags behind OpenAI in Top 1 by a margin of 5\%, but it still achieves better performance in MAP and MRR by 10\% and 4\%, respectively.
It demonstrates enhancements over SemanticCodeBERT, improving by 23\% in both Top 1 and MAP, and 18\% in MRR. 
Similarly, against FBL-BERT, \bz achieves substantial performance increases of 60\% in Top 1, 50\% in MAP, and 44\% in MRR.\par

When analyzing the performance of \bz across the studied programming languages, we find that \bz consistently outperforms the studied models in Top-1 scores across most programming languages.
The exception is Go, where OpenAI slightly outperforms \bz in Go projects.
We find that Go projects contain only 758 bugs, the fewest in our dataset. 
Additionally, the unique concurrency model and error-handling patterns of Go may pose challenges for \bz and contribute to lower performance in this language.
Moreover, \bz surpasses OpenAI with improvements ranging from 14.24\% to 16.86\% in all languages except Go, where it lags by 0.66\%. 
Compared to OpenAI, \bz achieves the highest improvement in Python and the lowest improvement in C$++$. Compared to SemanticCodeBERT, \bz shows consistent improvement between 15\% and 41\%, with the highest being observed in JavaScript and the lowest in Go. 
The trend continues when compared with FBL-BERT, where improvements range from 34.1\% to 67.5\%, peaking in Java and being lowest in JavaScript.




\section{Ablation Study}
\label{sec: ablation study}

This section presents an ablation study to assess the influence of individual components in the \bz pipeline on its overall performance. Key strategies contributing to the effectiveness of \bz include (1) a combined pipeline of lexical and deep retrievers, (2) learning from hard examples, and (3) dynamic chunking. For all the experiments in this ablation study, we use \bbox as it contains bugs from five different languages. Next, we detail the results of the ablation study to evaluate the impact of each component.


\subsection{Impact of hybrid retrieval technique}

\begin{figure}[tb!]
  \centering
  \begin{tcolorbox}[colback=white, boxrule=0.8pt, sharp corners, boxsep=0.5pt, left=0.5pt, right=0.5pt, top=0.5pt, bottom=0.5pt]
    \includegraphics[scale=0.165]{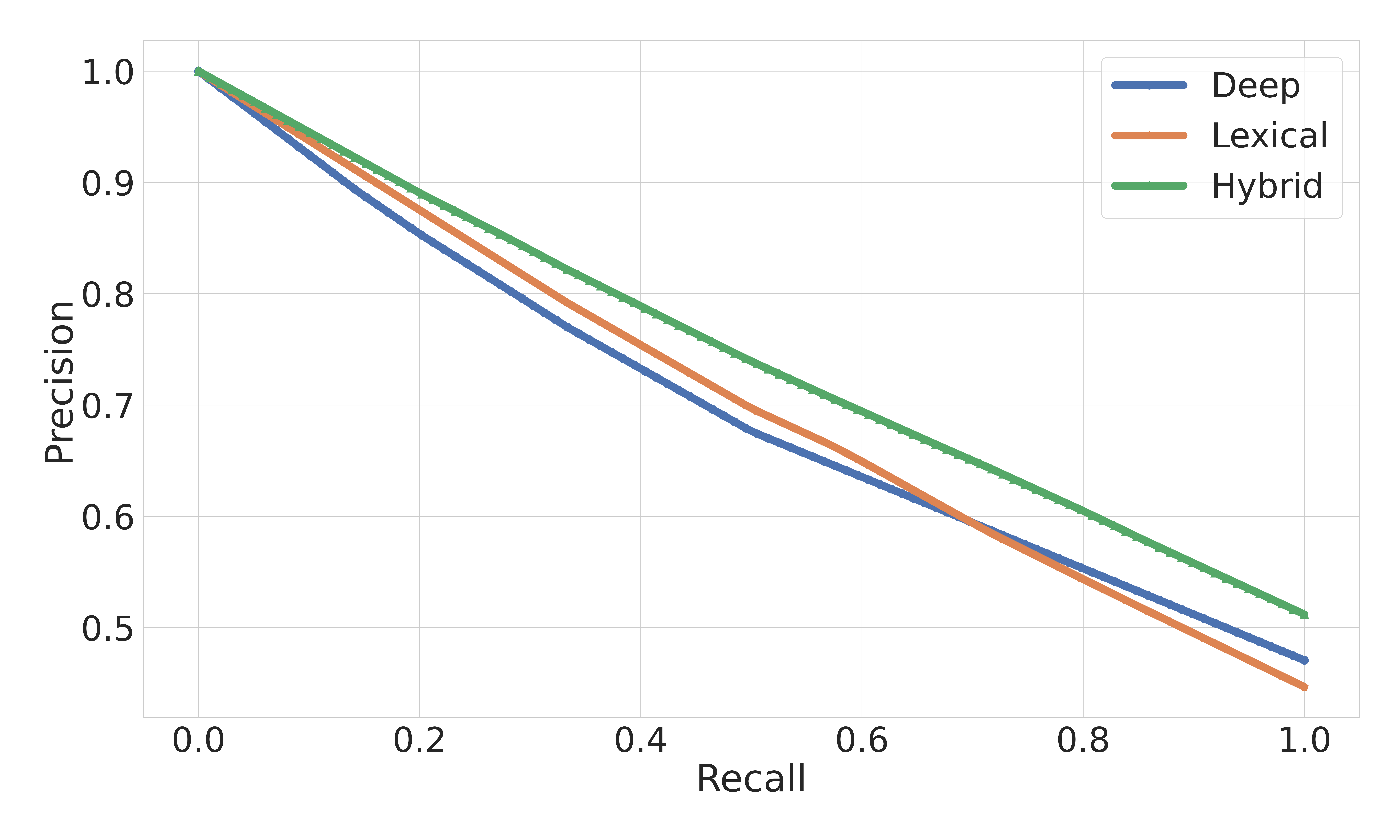}
  \end{tcolorbox}
   \caption{Precision recall curves of the retrievers.}
    \label{fig:pre_curve}
\end{figure}
\begin{table}[tb!]
\centering
\caption{Performance comparison between Lexical, Deep, and Hybrid Retriever.}
\resizebox{0.78\columnwidth}{!}{%
\begin{tabular}{@{}lrrrrr@{}}
\toprule
\textbf{Retriver Type} & \textbf{Top 1} & \textbf{Top 5} & \textbf{Top 10} & \textbf{MAP} & \textbf{MRR} \\ \midrule
Lexical                & 0.07           & 0.12           & 0.19            & 0.11         & 0.13         \\ \midrule
Deep                    & 0.12           & 0.25           & 0.31            & 0.14         & 0.18         \\ \midrule
Hybrid (\bz)          & \textbf{0.18}           & \textbf{0.37}           & \textbf{0.45}            & \textbf{0.22}         & \textbf{0.27}         \\ \bottomrule
\end{tabular}%
}
\label{table:retriever_ablation}
\end{table}


In \bz, we use a hybrid technique by combining outputs from two types of retrievers: lexical and deep. 
The lexical retriever is typically useful at finding exact matches and managing simple searches, while the deep retriever is better suited for understanding complex queries through their semantics. 
We assess their combined performance with a precision-recall graph, as shown in Figure~\ref{fig:pre_curve}. 
The figure demonstrates how accurately \bz identifies buggy source code files at various decision thresholds, with higher precision and recall indicating better performance.

Figure~\ref{fig:pre_curve} shows that the lexical retriever initially has high precision but low recall, indicating it identifies relevant files well but may miss others. 
As it retrieves more files, precision drops. Conversely, the deep retriever's precision increases as it processes more files, particularly at higher recall levels. 
By combining both retrievers in the \bz pipeline (as denoted `Hybrid' in Figure~\ref{fig:pre_curve}), we achieve a balance that enhances overall results.


Furthermore, we analyze the performance of each retriever and their combined effect in the \bz pipeline on the \bbox dataset. From Table~\ref{table:retriever_ablation}, we observe that the deep retriever outperforms the lexical retriever, showing improvements of 71\% in Top 1, 27\% in MAP, and 38\% in MRR. 
Combining the retrievers further enhances performance, showing improvements of 45\% to 57\% over the deep retriever alone. 
This result demonstrates that combining both retriever types mitigates their individual weaknesses and enhances the overall performance of \bz.


\subsection{Impact of fine-tuning}

\begin{figure*}[tb!]
  \centering
  \begin{subfigure}[t]{0.32\textwidth}
    \centering
    \fbox{
    \includegraphics[width=0.85\textwidth]{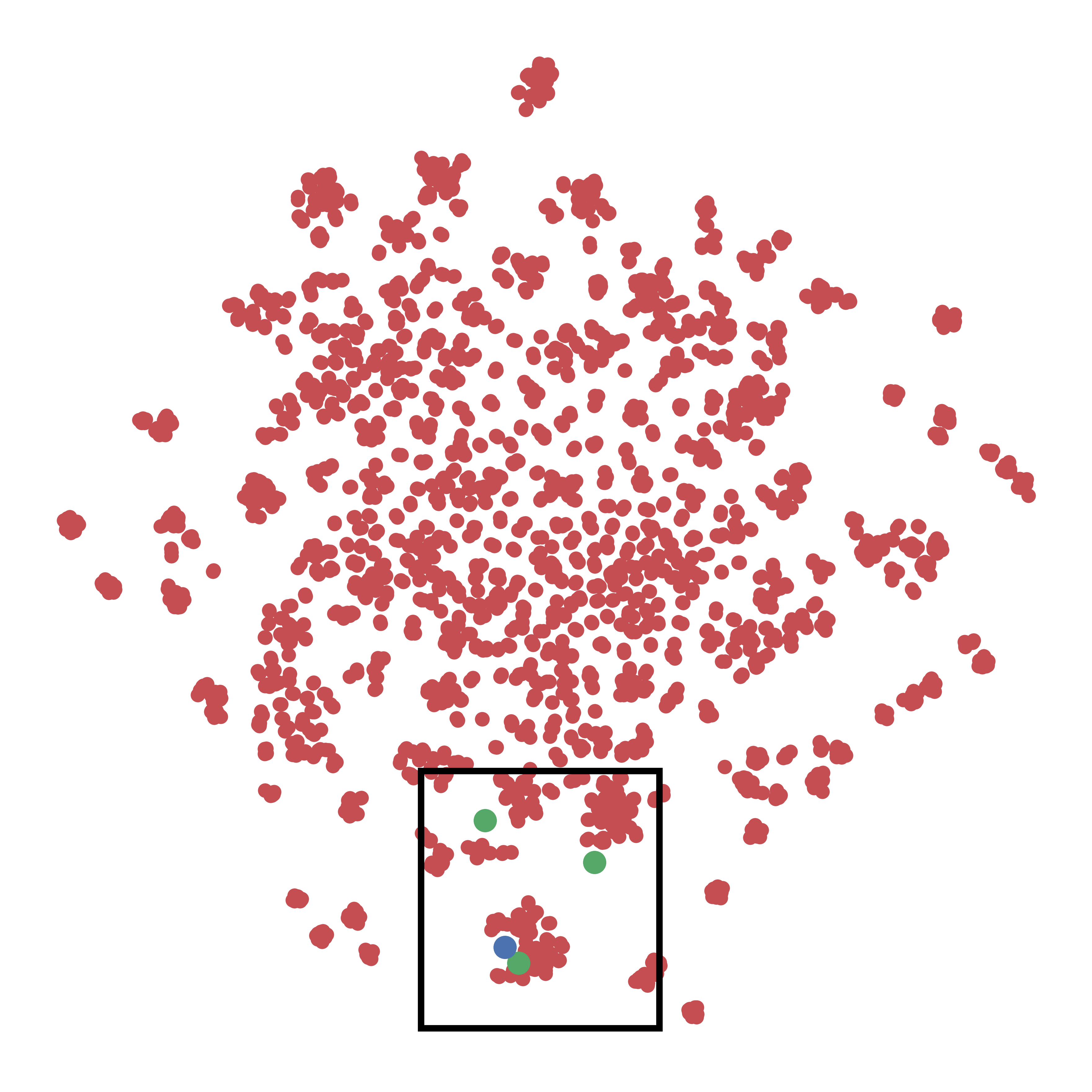}
    }
    \caption{Epoch 0 (CodeSage)}
    \label{fig:cluster_0}
  \end{subfigure}\quad
    \begin{subfigure}[t]{0.32\textwidth}
    \centering
    \fbox{
    \includegraphics[width=0.85\textwidth]{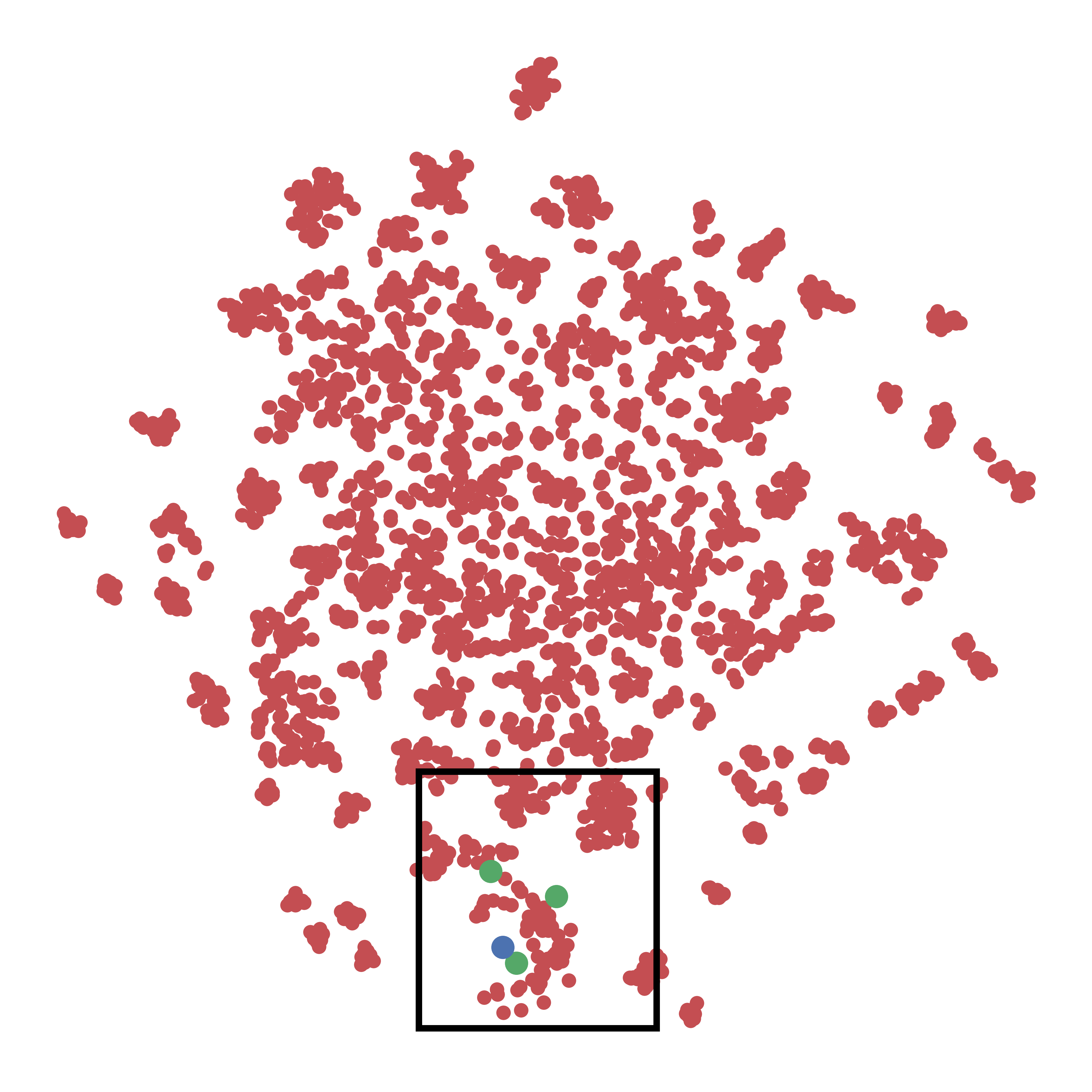}
    }
    \caption{Epoch 5}
    \label{fig:cluster_5}
  \end{subfigure}\quad
    \begin{subfigure}[t]{0.32\textwidth}
    \centering
    \fbox{
    \includegraphics[width=0.85\textwidth]{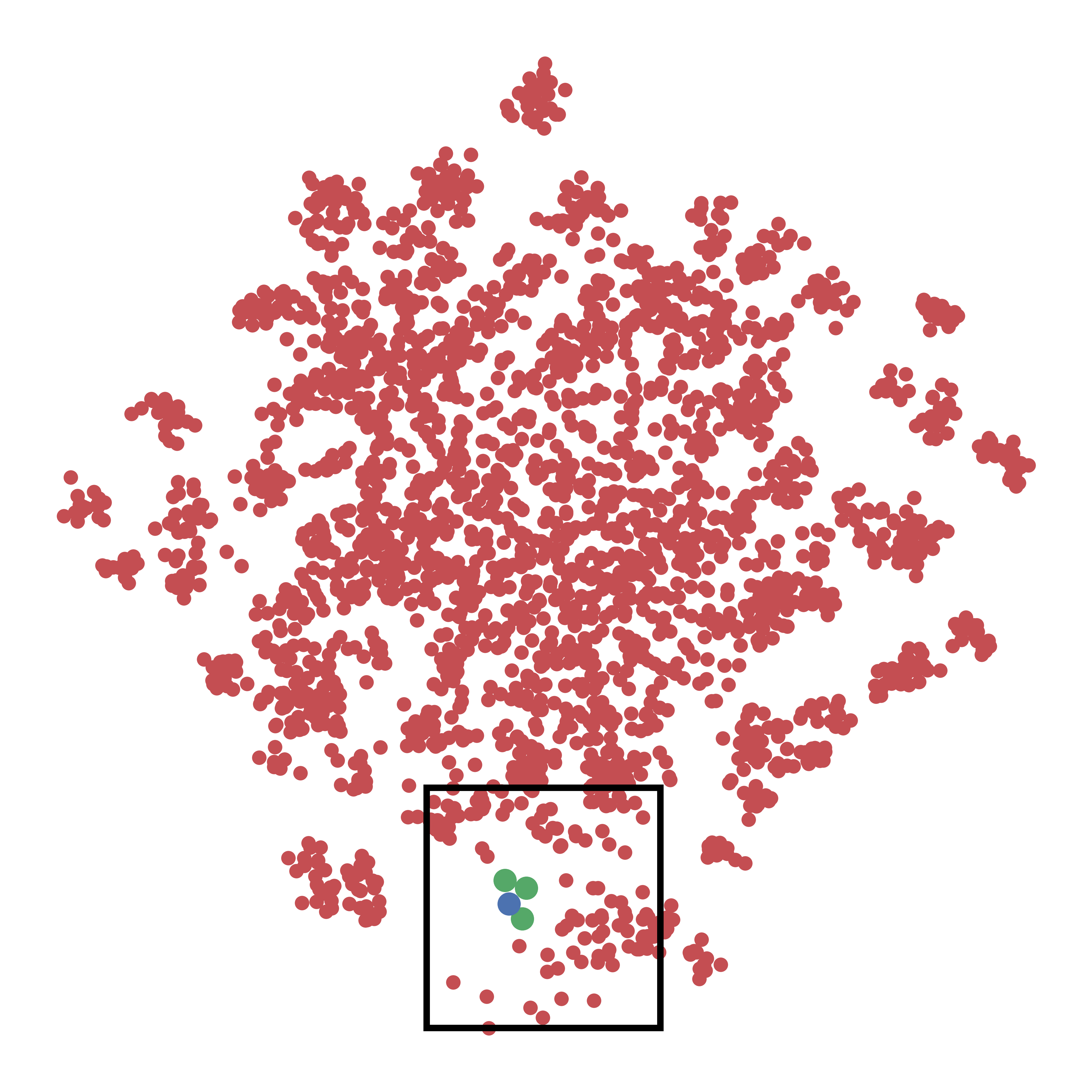}
    }
    \caption{Epoch 15 (\bz)}
    \label{fig:cluster_15}
  \end{subfigure}\quad
   \caption{{Scatter plots showing the chunks of the buggy and non-buggy files associated with the Apache Dubbo-1216 bug report. \raisebox{-0.4ex}{\textcolor{plotBlue}{\scalebox{2}{\textbullet}}} denotes bug report,  \raisebox{-0.4ex}{\textcolor{plotGreen}{\scalebox{2}{\textbullet}}} denotes chunks from buggy files, and \raisebox{-0.4ex}{\textcolor{plotRed}{\scalebox{2}{\textbullet}}} denotes chunks from non-buggy files.}}
\label{fig:cluster}
\end{figure*}

In \bz, we fine-tune a GPT-based model~\cite{zhang2024codesage} to better target hard-to-localize bugs by adjusting how closely buggy (positive files) and non-buggy (negative files) source code relate to bug reports (see~\ref{subsec: train_methodology}). 
We assess the impact of this fine-tuning by comparing the performance of the model at different stages: epoch 0 (CodeSage), epoch 5, and epoch 15 (\bz). We use t-SNE~\cite{van2008visualizing}, a method that simplifies complex data into two or three dimensions for easier visualization, to generate and display embeddings for a specific bug report (i.e., \emph{Apache Dubbo-1216}) and it is corresponding source code.
%

Figure~\ref{fig:cluster} shows that the ability of the CodeSage model to distinguish between buggy and non-buggy files in the analyzed bug report (i.e., \emph{Apache Dubbo-1216)} evolves over time. 
At the start (epoch 0), the model struggles to differentiate, with overlapping positive and negative chunks. 
By epoch 5, there is a clear improvement: positive chunks start clustering more closely, and negative chunks spread out, although some overlap remains. 
By epoch 15, the model has made substantial progress; positive chunks are well-grouped, and no negative chunks are closer than the positives, indicating successful fine-tuning.
\begin{table}[tb!]
\centering
\caption{Performance of CodeSage vs. \bz.}
\resizebox{0.78\columnwidth}{!}{%
\begin{tabular}{@{}lrrrrr@{}}
\toprule
\textbf{Retriver Type} & \textbf{Top 1} & \textbf{Top 5} & \textbf{Top 10} & \textbf{MAP} & \textbf{MRR} \\ \midrule
CodeSage\textsubscript{Deep}               & 0.11           & 0.21           & 0.28            & 0.12         & 0.16         \\ 
CodeSage\textsubscript{Hybrid} & 0.17           & 0.32           & 0.42            & 0.19         & 0.24         \\ 
\bz\textsubscript{Deep}             & 0.12           & 0.25           & 0.31            & 0.14         & 0.18         \\ 
\bz\textsubscript{Hybrid} & \textbf{0.18}           & \textbf{0.37}           & \textbf{0.45}            & \textbf{0.22}         & \textbf{0.27}         \\ \bottomrule
\end{tabular}%
}
\label{table:ft_ablation}
\end{table}

Also, we evaluate the CodeSage model on the \bbox dataset.
Table~\ref{table:ft_ablation}~shows the results of the CodeSage and \bz model with different types of retrievers.  For instance, CodeSage\textsubscript{Deep} reflects the performance with a deep retriever, while CodeSage\textsubscript{Hybrid} indicates the results using a hybrid retriever combining deep and lexical retriever. Using only the deep retriever, \bz achieves a performance improvement of 6.9\%--12.5\% over CodeSage, including 9.1\% for Top1, 16.7\% for MAP, and 12.5\% for MRR.
The hybrid technique used by the \bz pipeline leads to even more enhancements, ranging from 5.8\% to 15.7\%, with gains of 5.8\% in Top1, 15.8\% in MAP, and 12.5\% in MRR.
These results support that fine-tuning the model on challenging (hard) examples improves its effectiveness.



\subsection{Impact of dynamic chunking}
In \bz, we use dynamic chunking to manage the issue of limited context size while reducing continuity loss. Previous research~\cite{ratner2023, ivgi2023} has used methods like sliding windows and static chunking. The sliding window method overlaps chunks to maintain continuity but complicates similarity calculations due to these overlaps. Static chunking, on the other hand, avoids overlaps but can lead to a loss of continuity. Our approach uses dynamic programming to find breakpoints that reduce continuity loss without shrinking the window size. We trained and evaluated \bz using all three chunking methods to see the effects of each.

\begin{table}[tb!]
\centering
\caption{Static vs. Sliding window vs. Dynamic chunking.}
\resizebox{0.8\columnwidth}{!}{%
\begin{tabular}{@{}lrrrrr@{}}
\toprule
\textbf{Model} & \textbf{Top 1} & \textbf{Top 5} & \textbf{Top 10} & \textbf{MAP} & \textbf{MRR} \\ \midrule
\bz\textsubscript{Static + Deep}     & 0.11           & 0.22           & 0.26            & 0.13         & 0.17         \\ \midrule
\bz\textsubscript{Sliding + Deep}    & 0.11           & 0.24           & 0.3             & 0.13         & 0.17         \\ \midrule
\bz\textsubscript{Dynamic + Deep}          & 0.12           & 0.25           & 0.31            & 0.14         & 0.18         \\ \midrule
\bz\textsubscript{Static + Hybrid}  & 0.15           & 0.33           & 0.37            & 0.18         & 0.25         \\ \midrule
\bz\textsubscript{Sliding + Hybrid} & 0.17           & 0.36           & 0.42            & 0.21         & 0.25         \\ \midrule
\bz\textsubscript{Dynamic + Hybrid} & \textbf{0.18}           & \textbf{0.37}           & \textbf{0.45}            & \textbf{0.22}         & \textbf{0.27}         \\ \bottomrule
\end{tabular}%
}
\label{table:chunking_ablation}
\end{table}

Table~\ref{table:chunking_ablation} presents a performance comparison of various chunking techniques. For instance, \bz\textsubscript{Static + Deep} represents the performance of static chunking paired with the deep retriever. The table shows that \bz\textsubscript{Dynamic + Deep} improves performance by 5.5--16\% over static chunking and by 3.23--8.33\% over the sliding window method. Within the hybrid technique of \bz pipeline, dynamic chunking shows a performance increase of 7--18\% compared to static chunking and 2--7.41\% compared to sliding window.


\begin{table}[tb!]
\centering
\caption{Resource usage analysis of Static, Sliding Window, and Dynamic Chunking.}
\label{table:resource_ablation}

\begin{tabular}{@{}lcc@{}}
\toprule
\textbf{Technique} & \multicolumn{1}{l}{\textbf{\# Chunks}} & \multicolumn{1}{l}{\textbf{TFLOPS}} \\ \midrule
Static             & 200,984                                & 380,864                             \\ \midrule
Sliding Window     & 274,260                                & 519,722                             \\ \midrule
Dynamic            & 209,359                                & 396,735                             \\ \bottomrule
\end{tabular}

\end{table}

We also analyzed the impact of chunking techniques on resource usage, measuring the number of chunks generated and TFLOPS (an indicator of computational cost)~\cite{Desislavov2023}. To measure TFLOPS, we utilized the built-in method provided by the PyTorch framework, which calculates TFLOPS based on the data size, model size, and model architecture. Table~\ref{table:resource_ablation} reveals that static chunking produces the fewest chunks and consumes the least TFLOPS while sliding windows generate the most chunks and require the highest computational resources. However, Table~\ref{table:chunking_ablation} indicates that despite its low resource consumption, static chunking performs poorly in bug localization tasks. Sliding windows, although resource-intensive, do not achieve optimal performance. In contrast, dynamic chunking offers a balance, achieving 2--7.41\% better performance than sliding windows while consuming 3\% fewer resources.

\section{Threats to Validity}
\label{sec:threats}
Below, we discuss the threats to the validity of our study.

\noindent
\textbf{Internal validity.}
One threat to internal validity is the hyperparameters we used to train the models. 
To mitigate this, we use the same parameters used by several prior studies that use large language models~\cite{devlin2019, Ciborowska2022, Khattab2020, Kochhar2014}. Although we applied conservative repository-level filtering, we recognize that some residual risk of leakage could persist due to indirect or partial content overlaps. However, given the lack of transparency in LLM training corpora, our approach represents a reasonable trade-off between feasibility and rigor. Furthermore, during the filtering process, we exclude bug reports that explicitly mention modified files. However, subtle or indirect references may still go undetected. This presents a residual threat to internal validity, as a small number of answer-revealing reports may persist despite our conservative heuristics.

\noindent
\textbf{External validity.}
A threat to the external validity of our study lies in the generalizability of the results. 
We assess cross-project capabilities using separate projects for training and testing. 
Although the choice of test projects may affect performance, this approach aligns with previous studies that evaluate zero-shot and cross-project capabilities~\cite{Fan2024, jimenez2024swebench}.
Additionally, our evaluation of 30,625 real-world bugs ensures that our findings hold, even with different project selections, due to the large scale of our dataset. 
To address potential selection bias in the \bbox dataset, we collected project repositories based on star ratings and recent updates, ensuring that the repositories are actively contributed to and popular within the open-source community. 
This approach aims to include high-quality, widely used projects across the selected top five programming languages. Additionally, due to the absence of a replication package, we replicated TRANP-CNN and CooBA based on their descriptions in the original study, which may lead to slight performance deviations. Nevertheless, after experimenting with various hyperparameters, we selected a set that achieves comparable performance to that reported in the original study.

\noindent
\textbf{Construct validity.}
Our selected evaluation metrics can threaten the construct validity of the study. 
We used Top N, MAP, and
MRR as evaluation metrics. 
These are well-known evaluation measures adopted in prior studies which are widely used when evaluating the performance of bug localization tools that rely on information retrieval techniques~\cite{Ciborowska2022, Liang2022, Ye2014, Zhou2012}. 

\section{Related Works}
\vspace{2mm}
\label{sec:related_works}
In this section, we present prior work related to bug localization approaches and discuss how our approach differs from them.

A plethora of work proposed lexical retriever-based approaches to localizing bugs.
Such approaches use BM25 and TF-IDF techniques to measure the similarity between a bug report and \scf~\cite{Zhou2012, Wang2014, Murali2021, Razzaq2021, Niu2023, almhana2021method}.
For example, BugLocator~\cite{Zhou2012} used TF-IDF to measure the similarity between the bug report and \scf using a revised vector space model. Sisman et al.~\cite{Sisman2012} and Amalgam~\cite{Wang2014} used version history along with the VSM model to localize bugs. 
Shi et al.~\cite{Shi2014} found that BM25 and its variants are better than TF-IDF, VSM, and rVSM in the bug localization task. BoostNSift~\cite{Razzaq2021} used BM25 for method-level bug localization. Lee et al.~\cite{Lee2018} proposed the Bench4BL dataset and evaluated six bug-localization tools. Our study differs from Lee et al.'s, mainly in the reliability of ground truth mappings. While Lee et al. used the Bench4BL dataset—later shown by Kim et al.~\cite{Kim2021} to have up to 44\% incorrect mappings—we applied Kim et al.’s method to ensure accuracy, resulting in only 0.06\% incorrect mappings. This improvement is likely due to our conservative issue selection, including only issues confirmed to be merged, unlike in Lee et al.’s approach.
Our approach, \bz, differs from prior work since it leverages the advantages of both the BM25-based retrieval method and deep learning-based technique to perform cross-project bug localization.

Other work proposed deep learning-based approaches to localizing bugs~\cite{Liang2022, Huo2016, Zhou2012, Zhu2021, Zhang2020, Qi2022, Chakraborty2024}.
For example, FLIM~\cite{Liang2022} used CodeBERT embedding for calculating the similarity between bug reports and \scf at the function level.
DNNLoc~\cite{Lam2017} combined a deep neural network with the VSM model to increase effectiveness across different types of metrics such as textual similarity, class name similarity, bug-fixing recency, and frequency. 
NP-CNN~\cite{Huo2016} employed a Convolutional Neural Network (CNN) to leverage both lexical and semantical properties. 
BL-GAN~\cite{Zhu2022} proposed a generative adversarial network-based solution for bug localization.
Several studies~\cite{Xiao2018, Xiao2019, Li2019, Liang2019} incorporated program structure information in the model to generate more useful program representation. 
For example, Liang et al.~\cite{Liang2019} and  Li et al.~\cite{Li2019} used abstract syntax trees, data flow graphs, and program flow graphs to generate a better representation of source code.
While \bz employs an embedding model similar to the techniques discussed previously, it differs by its ability to be applied to new, unseen projects without requiring additional training.

Several studies offered test case coverage or spectrum-based solutions for bug localization~\cite{Vancsics2022, Kim2019, Wen2021, Widyasari2022, Yang2024, Zhang2023, Ghanbari2023, Lucia2014}. For example, FusionLocalizer~\cite{Lucia2014} combines multiple test case-based fault localizers. 
However, their approach depends on tests, whereas \bz uses an information retrieval approach solely based on bug reports. 
Moreover, the selection step in the FusionLocalizer technique may introduce additional bias, while \bz employs Reciprocal Rank Fusion (RRF) to consider the ranks of all retrievers, effectively reducing bias. Similarly, UniLoc~\cite{Hassan2023} uses information retrieval for localizing bugs in continuous integration tools, while \bz focuses on bug reports.
Additionally, CrossLocator~\cite{Xia2014} introduces a translation-based method for projects that uses multiple languages for bug reports (e.g., English, Chinese), whereas \bz identifies bugs in projects involving different programming languages.


While numerous studies have been conducted on bug localization tools, most of them are project-context approaches, i.e., the research on cross-project bug localization is limited in comparison.
For example, TRANP-CNN~\cite{Huo2021} used a combination of transfer learning and CNN models to attain competitive performance in cross-project bug localization at the file level.
CooBa~\cite{Zhu2020} follows a similar approach but employs a shared encoder to capture cross-project features.
Such approaches still require further training before deploying and making them work for new projects. 
In contrast, \bz is designed in a way that it can work on a new project without further training. DSSDPP~\cite{Li2023} investigated a sampling approach to address disparities between source and target projects, whereas xLoc~\cite{yangh2024} used transformers to identify bugs in cross-language API calls. Nevertheless, both tools concentrate on method-level defect prediction, in contrast to \bz, which localizes bugs at the file level.

Previous research has also explored embedding-based methods for bug localization. SemanticCodeBERT~\cite{Du2023} integrates semantic flow graph data, while FBL-BERT~\cite{Ciborowska2022} uses changeset data. However, neither focuses on hard examples to improve feature separation. In contrast, \bz learns from hard examples to significantly enhance model performance.
\vspace{2mm}

\section{Conclusion}
\label{sec:conclusion}
In this paper, we introduce \bz, a bug localization tool that operates without project-specific training, facilitating cross-project and cross-language bug localization. 
\bz employs a dynamic chunking and hard example learning methodology to model lengthy source code files. 
To support cross-project and cross-language bug localizers, we curate \bbox---one of the largest (if not \emph{the} largest) cross-project and cross-language bug localization datasets.
We evaluate \bz with 30,625 bug reports from 34 real-world projects, showing up to a 100\% improvement in the Top 1 metric against cross-project localizers and up to a 60\% improvement over modern large language model approaches. Our results suggest that the use of dynamic chunking and hard example learning is key in enhancing the performance of bug localization tools.

\bibliographystyle{IEEEtran}
\bibliography{Bibliography}

\begin{thebibliography}{10}
\providecommand{\url}[1]{#1}
\csname url@samestyle\endcsname
\providecommand{\newblock}{\relax}
\providecommand{\bibinfo}[2]{#2}
\providecommand{\BIBentrySTDinterwordspacing}{\spaceskip=0pt\relax}
\providecommand{\BIBentryALTinterwordstretchfactor}{4}
\providecommand{\BIBentryALTinterwordspacing}{\spaceskip=\fontdimen2\font plus
\BIBentryALTinterwordstretchfactor\fontdimen3\font minus
  \fontdimen4\font\relax}
\providecommand{\BIBforeignlanguage}[2]{{%
\expandafter\ifx\csname l@#1\endcsname\relax
\typeout{** WARNING: IEEEtran.bst: No hyphenation pattern has been}%
\typeout{** loaded for the language `#1'. Using the pattern for}%
\typeout{** the default language instead.}%
\else
\language=\csname l@#1\endcsname
\fi
#2}}
\providecommand{\BIBdecl}{\relax}
\BIBdecl

\bibitem{Koomen1999}
T.~Koomen and M.~Pol, \emph{Test Process Improvement: A Practical Step-by-Step
  Guide to Structured Testing}.\hskip 1em plus 0.5em minus 0.4em\relax USA:
  Addison-Wesley Longman Publishing Co., Inc., 1999.

\bibitem{LaToza2010}
T.~D. LaToza and B.~A. Myers, ``Developers ask reachability questions,'' in
  \emph{Proceedings of the 32nd {ACM}/{IEEE} International Conference on
  Software Engineering - {ICSE} {\textquotesingle}10}.\hskip 1em plus 0.5em
  minus 0.4em\relax {ACM} Press, 2010.

\bibitem{Tassey2002}
G.~Tassey, ``The economic impacts of inadequate infrastructure for software
  testing,'' 2002.

\bibitem{ElDeeb2023}
A.~El-Deeb, ``What the software industry is measuring?'' \emph{{ACM} {SIGSOFT}
  Software Engineering Notes}, vol.~48, no.~2, pp. 10--11, Apr. 2023.

\bibitem{Zhou2012}
J.~Zhou, H.~Zhang, and D.~Lo, ``Where should the bugs be fixed? more accurate
  information retrieval-based bug localization based on bug reports,'' in
  \emph{2012 34th International Conference on Software Engineering
  ({ICSE})}.\hskip 1em plus 0.5em minus 0.4em\relax {IEEE}, Jun. 2012.

\bibitem{Wang2014}
S.~Wang and D.~Lo, ``Version history, similar report, and structure: putting
  them together for improved bug localization,'' in \emph{Proceedings of the
  22nd International Conference on Program Comprehension}.\hskip 1em plus 0.5em
  minus 0.4em\relax {ACM}, Jun. 2014.

\bibitem{Fischer2003}
M.~Fischer, M.~Pinzger, and H.~Gall, ``Analyzing and relating bug report data
  for feature tracking,'' in \emph{10th Working Conference on Reverse
  Engineering, 2003. {WCRE} 2003. Proceedings.}\hskip 1em plus 0.5em minus
  0.4em\relax {IEEE}, 2003.

\bibitem{Li2019}
Y.~Li, S.~Wang, T.~N. Nguyen, and S.~V. Nguyen, ``Improving bug detection via
  context-based code representation learning and attention-based neural
  networks,'' \emph{Proceedings of the {ACM} on Programming Languages}, vol.~3,
  no. {OOPSLA}, pp. 1--30, Oct. 2019.

\bibitem{Xiao2019}
Y.~Xiao, J.~Keung, K.~E. Bennin, and Q.~Mi, ``Improving bug localization with
  word embedding and enhanced convolutional neural networks,''
  \emph{Information and Software Technology}, vol. 105, pp. 17--29, Jan. 2019.

\bibitem{Liang2019}
H.~Liang, L.~Sun, M.~Wang, and Y.~Yang, ``Deep learning with customized
  abstract syntax tree for bug localization,'' \emph{{IEEE} Access}, vol.~7,
  pp. 116\,309--116\,320, 2019.

\bibitem{Zhang2020}
J.~Zhang, R.~Xie, W.~Ye, Y.~Zhang, and S.~Zhang, ``Exploiting code knowledge
  graph for bug localization via bi-directional attention,'' in
  \emph{Proceedings of the 28th International Conference on Program
  Comprehension}.\hskip 1em plus 0.5em minus 0.4em\relax {ACM}, Jul. 2020.

\bibitem{Zhu2022}
Z.~Zhu, H.~Tong, Y.~Wang, and Y.~Li, ``{BL}-{GAN}: Semi-supervised bug
  localization via generative adversarial network,'' \emph{{IEEE} Transactions
  on Knowledge and Data Engineering}, pp. 1--14, 2022.

\bibitem{Kochhar2014}
P.~S. Kochhar, Y.~Tian, and D.~Lo, ``Potential biases in bug localization,'' in
  \emph{Proceedings of the 29th {ACM}/{IEEE} International Conference on
  Automated Software Engineering}.\hskip 1em plus 0.5em minus 0.4em\relax
  {ACM}, Sep. 2014.

\bibitem{Sainani2020}
A.~Sainani, P.~R. Anish, V.~Joshi, and S.~Ghaisas, ``Extracting and classifying
  requirements from software engineering contracts,'' in \emph{2020 {IEEE} 28th
  International Requirements Engineering Conference ({RE})}.\hskip 1em plus
  0.5em minus 0.4em\relax {IEEE}, Aug. 2020.

\bibitem{Huo2021}
X.~Huo, F.~Thung, M.~Li, D.~Lo, and S.-T. Shi, ``Deep transfer bug
  localization,'' \emph{{IEEE} Transactions on Software Engineering}, vol.~47,
  no.~7, pp. 1368--1380, Jul. 2021.

\bibitem{Zhu2020}
Z.~Zhu, Y.~Li, H.~Tong, and Y.~Wang, ``{CooBa}: Cross-project bug localization
  via adversarial transfer learning,'' in \emph{Proceedings of the Twenty-Ninth
  International Joint Conference on Artificial Intelligence}.\hskip 1em plus
  0.5em minus 0.4em\relax International Joint Conferences on Artificial
  Intelligence Organization, Jul. 2020.

\bibitem{feng2020}
Z.~Feng, D.~Guo, D.~Tang, N.~Duan, X.~Feng, M.~Gong, L.~Shou, B.~Qin, T.~Liu,
  D.~Jiang, and M.~Zhou, ``{C}ode{BERT}: A pre-trained model for programming
  and natural languages,'' in \emph{Findings of the Association for
  Computational Linguistics: EMNLP 2020}.\hskip 1em plus 0.5em minus
  0.4em\relax Online: Association for Computational Linguistics, Nov. 2020, pp.
  1536--1547.

\bibitem{guo2021}
D.~Guo, S.~Ren, S.~Lu, Z.~Feng, D.~Tang, S.~Liu, L.~Zhou, N.~Duan,
  A.~Svyatkovskiy, S.~Fu, M.~Tufano, S.~K. Deng, C.~Clement, D.~Drain,
  N.~Sundaresan, J.~Yin, D.~Jiang, and M.~Zhou, ``Graphcodebert: Pre-training
  code representations with data flow,'' 2021.

\bibitem{Ciborowska2022}
A.~Ciborowska and K.~Damevski, ``Fast changeset-based bug localization with
  {BERT},'' in \emph{Proceedings of the 44th International Conference on
  Software Engineering}.\hskip 1em plus 0.5em minus 0.4em\relax {ACM}, May
  2022.

\bibitem{Du2023}
Y.~Du and Z.~Yu, ``Pre-training code representation with semantic flow graph
  for effective bug localization,'' in \emph{Proceedings of the 31th ACM Joint
  Meeting on European Software Engineering Conference and Symposium on the
  Foundations of Software Engineering}, ser. ESEC/FSE 2023, 2023.

\bibitem{Ye2014}
X.~Ye, R.~Bunescu, and C.~Liu, ``Learning to rank relevant files for bug
  reports using domain knowledge,'' in \emph{Proceedings of the 22nd {ACM}
  {SIGSOFT} International Symposium on Foundations of Software Engineering -
  {FSE} 2014}.\hskip 1em plus 0.5em minus 0.4em\relax {ACM} Press, 2014.

\bibitem{qiao2023}
Z.~Qiao, W.~Ye, D.~Yu, T.~Mo, W.~Li, and S.~Zhang, ``Improving knowledge graph
  completion with generative hard negative mining,'' in \emph{Findings of the
  Association for Computational Linguistics: ACL 2023}, A.~Rogers,
  J.~Boyd-Graber, and N.~Okazaki, Eds.\hskip 1em plus 0.5em minus 0.4em\relax
  Toronto, Canada: Association for Computational Linguistics, Jul. 2023, pp.
  5866--5878.

\bibitem{Xuan2020}
H.~Xuan, A.~Stylianou, X.~Liu, and R.~Pless, \emph{Hard Negative Examples are
  Hard, but Useful}.\hskip 1em plus 0.5em minus 0.4em\relax Springer
  International Publishing, 2020, p. 126–142.

\bibitem{Ray2017}
B.~Ray, D.~Posnett, P.~Devanbu, and V.~Filkov, ``A large-scale study of
  programming languages and code quality in github,'' \emph{Communications of
  the ACM}, vol.~60, no.~10, p. 91–100, Sep. 2017.

\bibitem{Zeller2002}
A.~Zeller and R.~Hildebrandt, ``Simplifying and isolating failure-inducing
  input,'' \emph{IEEE Transactions on Software Engineering}, vol.~28, no.~2, p.
  183–200, 2002.

\bibitem{liwerski2005}
J.~Śliwerski, T.~Zimmermann, and A.~Zeller, ``When do changes induce fixes?''
  \emph{ACM SIGSOFT Software Engineering Notes}, vol.~30, no.~4, p. 1–5, May
  2005.

\bibitem{Canfora2005}
G.~Canfora and L.~Cerulo, ``Impact analysis by mining software and change
  request repositories,'' in \emph{11th IEEE International Software Metrics
  Symposium (METRICS’05)}.\hskip 1em plus 0.5em minus 0.4em\relax IEEE, 2005.

\bibitem{Hassan2004}
A.~Hassan and R.~Holt, ``Predicting change propagation in software systems,''
  in \emph{20th IEEE International Conference on Software Maintenance, 2004.
  Proceedings.}\hskip 1em plus 0.5em minus 0.4em\relax IEEE, 2004.

\bibitem{Rahman2013}
F.~Rahman and P.~Devanbu, ``How, and why, process metrics are better,'' in
  \emph{2013 35th International Conference on Software Engineering
  (ICSE)}.\hskip 1em plus 0.5em minus 0.4em\relax IEEE, May 2013.

\bibitem{Herzig2013}
K.~Herzig, S.~Just, and A.~Zeller, ``It{\textquotesingle}s not a bug,
  it{\textquotesingle}s a feature: How misclassification impacts bug
  prediction,'' in \emph{2013 35th International Conference on Software
  Engineering ({ICSE})}.\hskip 1em plus 0.5em minus 0.4em\relax {IEEE}, May
  2013.

\bibitem{Zimmermann2007}
T.~Zimmermann, R.~Premraj, and A.~Zeller, ``Predicting defects for eclipse,''
  in \emph{Third International Workshop on Predictor Models in Software
  Engineering (PROMISE’07: ICSE Workshops 2007)}.\hskip 1em plus 0.5em minus
  0.4em\relax IEEE, May 2007.

\bibitem{jimenez2024swebench}
C.~E. Jimenez, J.~Yang, A.~Wettig, S.~Yao, K.~Pei, O.~Press, and K.~R.
  Narasimhan, ``{SWE}-bench: Can language models resolve real-world github
  issues?'' in \emph{The Twelfth International Conference on Learning
  Representations}, 2024.

\bibitem{Neelakantan2022}
A.~Neelakantan, T.~Xu, R.~Puri, A.~Radford, J.~M. Han, J.~Tworek, Q.~Yuan,
  N.~Tezak, J.~W. Kim, C.~Hallacy, J.~Heidecke, P.~Shyam, B.~Power, T.~E.
  Nekoul, G.~Sastry, G.~Krueger, D.~Schnurr, F.~P. Such, K.~Hsu, M.~Thompson,
  T.~Khan, T.~Sherbakov, J.~Jang, P.~Welinder, and L.~Weng, ``Text and code
  embeddings by contrastive pre-training,'' 2022.

\bibitem{rep_package}
\BIBentryALTinterwordspacing
{Anonymous}, ``\BIBforeignlanguage{en}{Blaze: Bug localization in zero-shot
  environment},'' \emph{\BIBforeignlanguage{en}{Zenodo}}, 2023. [Online].
  Available: \url{https://zenodo.org/record/7897697}
\BIBentrySTDinterwordspacing

\bibitem{Liu2021}
C.~Liu, X.~Xia, D.~Lo, Z.~Liu, A.~E. Hassan, and S.~Li, ``{CodeMatcher}:
  Searching code based on sequential semantics of important query words,''
  \emph{{ACM} Transactions on Software Engineering and Methodology}, vol.~31,
  no.~1, pp. 1--37, Sep. 2021.

\bibitem{devlin2019}
J.~Devlin, M.-W. Chang, K.~Lee, and K.~Toutanova, ``{BERT}: Pre-training of
  deep bidirectional transformers for language understanding,'' in
  \emph{Proceedings of the 2019 Conference of the North {A}merican Chapter of
  the Association for Computational Linguistics: Human Language Technologies,
  Volume 1 (Long and Short Papers)}.\hskip 1em plus 0.5em minus 0.4em\relax
  Minneapolis, Minnesota: Association for Computational Linguistics, Jun. 2019,
  pp. 4171--4186.

\bibitem{Liu2019}
Y.~Liu, M.~Ott, N.~Goyal, J.~Du, M.~Joshi, D.~Chen, O.~Levy, M.~Lewis,
  L.~Zettlemoyer, and V.~Stoyanov, ``Roberta: A robustly optimized bert
  pretraining approach,'' 2019.

\bibitem{Beltagy2020}
I.~Beltagy, M.~E. Peters, and A.~Cohan, ``Longformer: The long-document
  transformer,'' 2020.

\bibitem{zhang2024codesage}
D.~Zhang*, W.~Ahmad*, M.~Tan, H.~Ding, R.~Nallapati, D.~Roth, X.~Ma, and
  B.~Xiang, ``Codesage: Code representation learning at scale,'' in \emph{The
  Twelfth International Conference on Learning Representations}, 2024.

\bibitem{Chen2020}
L.~Chen, Z.~Tang, and G.~H. Yang, ``Balancing reinforcement learning training
  experiences in interactive information retrieval,'' in \emph{Proceedings of
  the 43rd International {ACM} {SIGIR} Conference on Research and Development
  in Information Retrieval}.\hskip 1em plus 0.5em minus 0.4em\relax {ACM}, Jul.
  2020.

\bibitem{Wang2021}
J.~Wang, X.~Yi, R.~Guo, H.~Jin, P.~Xu, S.~Li, X.~Wang, X.~Guo, C.~Li, X.~Xu,
  K.~Yu, Y.~Yuan, Y.~Zou, J.~Long, Y.~Cai, Z.~Li, Z.~Zhang, Y.~Mo, J.~Gu,
  R.~Jiang, Y.~Wei, and C.~Xie, ``Milvus: A purpose-built vector data
  management system,'' in \emph{Proceedings of the 2021 International
  Conference on Management of Data}, ser. SIGMOD/PODS ’21.\hskip 1em plus
  0.5em minus 0.4em\relax ACM, Jun. 2021.

\bibitem{elasticsearch}
C.~Gormley and Z.~Tong, \emph{Elasticsearch: The Definitive Guide},
  1st~ed.\hskip 1em plus 0.5em minus 0.4em\relax O'Reilly Media, Inc., 2015.

\bibitem{bm25}
S.~Robertson, S.~Walker, S.~Jones, M.~M. Hancock-Beaulieu, and M.~Gatford,
  ``Okapi at trec-3,'' in \emph{Overview of the Third Text REtrieval Conference
  (TREC-3)}.\hskip 1em plus 0.5em minus 0.4em\relax Gaithersburg, MD: NIST,
  January 1995, pp. 109--126.

\bibitem{Sachdev2018}
S.~Sachdev, H.~Li, S.~Luan, S.~Kim, K.~Sen, and S.~Chandra, ``Retrieval on
  source code: a neural code search,'' in \emph{Proceedings of the 2nd {ACM}
  {SIGPLAN} International Workshop on Machine Learning and Programming
  Languages}.\hskip 1em plus 0.5em minus 0.4em\relax {ACM}, Jun. 2018.

\bibitem{Abbasiantaeb2021}
Z.~Abbasiantaeb and S.~Momtazi, ``Text-based question answering from
  information retrieval and deep neural network perspectives: A survey,''
  \emph{{WIREs} Data Mining and Knowledge Discovery}, vol.~11, no.~6, May 2021.

\bibitem{Luan2021}
Y.~Luan, J.~Eisenstein, K.~Toutanova, and M.~Collins, ``Sparse, dense, and
  attentional representations for text retrieval,'' \emph{Transactions of the
  Association for Computational Linguistics}, vol.~9, pp. 329--345, 2021.

\bibitem{Wangb2021}
S.~Wang, S.~Zhuang, and G.~Zuccon, ``{BERT}-based dense retrievers require
  interpolation with {BM}25 for effective passage retrieval,'' in
  \emph{Proceedings of the 2021 {ACM} {SIGIR} International Conference on
  Theory of Information Retrieval}.\hskip 1em plus 0.5em minus 0.4em\relax
  {ACM}, Jul. 2021.

\bibitem{Cormack2009}
G.~V. Cormack, C.~L.~A. Clarke, and S.~Buettcher, ``Reciprocal rank fusion
  outperforms condorcet and individual rank learning methods,'' in
  \emph{Proceedings of the 32nd international {ACM} {SIGIR} conference on
  Research and development in information retrieval}.\hskip 1em plus 0.5em
  minus 0.4em\relax {ACM}, jul 2009.

\bibitem{Bendersky2020}
M.~Bendersky, H.~Zhuang, J.~Ma, S.~Han, K.~Hall, and R.~McDonald, ``Rrf102:
  Meeting the trec-covid challenge with a 100+ runs ensemble,'' 2020.

\bibitem{Chen2022rrf}
T.~Chen, M.~Zhang, J.~Lu, M.~Bendersky, and M.~Najork, ``Out-of-domain
  semantics to~the~rescue! zero-shot hybrid retrieval models,'' in
  \emph{Lecture Notes in Computer Science}.\hskip 1em plus 0.5em minus
  0.4em\relax Springer International Publishing, 2022, pp. 95--110.

\bibitem{Xiao2018}
Y.~Xiao, J.~Keung, K.~E. Bennin, and Q.~Mi, ``Machine translation-based bug
  localization technique for bridging lexical gap,'' \emph{Information and
  Software Technology}, vol.~99, p. 58–61, Jul. 2018.

\bibitem{Huo2016}
X.~Huo, M.~Li, and Z.-H. Zhou, ``Learning unified features from natural and
  programming languages for locating buggy source code,'' in \emph{IJCAI},
  2016.

\bibitem{Moreno2014}
L.~Moreno, J.~J. Treadway, A.~Marcus, and W.~Shen, ``On the use of stack traces
  to improve text retrieval-based bug localization,'' in \emph{2014 IEEE
  International Conference on Software Maintenance and Evolution}.\hskip 1em
  plus 0.5em minus 0.4em\relax IEEE, Sep. 2014.

\bibitem{Dit2011}
B.~Dit, M.~Revelle, M.~Gethers, and D.~Poshyvanyk, ``Feature location in source
  code: a taxonomy and survey,'' \emph{Journal of Software: Evolution and
  Process}, vol.~25, no.~1, p. 53–95, Nov. 2011.

\bibitem{Lee2018}
J.~Lee, D.~Kim, T.~F. Bissyandé, W.~Jung, and Y.~Le~Traon, ``Bench4bl:
  reproducibility study on the performance of ir-based bug localization,'' in
  \emph{Proceedings of the 27th ACM SIGSOFT International Symposium on Software
  Testing and Analysis}, ser. ISSTA ’18.\hskip 1em plus 0.5em minus
  0.4em\relax ACM, Jul. 2018.

\bibitem{Kim2021}
M.~Kim and E.~Lee, ``Are datasets for information retrieval-based bug
  localization techniques trustworthy?: Impact analysis of bug types on irbl,''
  \emph{Empirical Software Engineering}, vol.~26, no.~3, Mar. 2021.

\bibitem{elangovan2021}
A.~Elangovan, J.~He, and K.~Verspoor, ``Memorization vs. generalization :
  Quantifying data leakage in {NLP} performance evaluation,'' in
  \emph{Proceedings of the 16th Conference of the European Chapter of the
  Association for Computational Linguistics: Main Volume}, P.~Merlo,
  J.~Tiedemann, and R.~Tsarfaty, Eds.\hskip 1em plus 0.5em minus 0.4em\relax
  Online: Association for Computational Linguistics, Apr. 2021, pp. 1325--1335.

\bibitem{Fan2024}
G.~Fan, S.~Chen, C.~Gao, J.~Xiao, T.~Zhang, and Z.~Feng, ``Rapid: Zero-shot
  domain adaptation for code search with pre-trained models,'' \emph{ACM
  Transactions on Software Engineering and Methodology}, vol.~33, no.~5, p.
  1–35, Jun. 2024.

\bibitem{van2008visualizing}
L.~Van~der Maaten and G.~Hinton, ``Visualizing data using t-sne.''
  \emph{Journal of machine learning research}, vol.~9, no.~11, 2008.

\bibitem{ratner2023}
N.~Ratner, Y.~Levine, Y.~Belinkov, O.~Ram, I.~Magar, O.~Abend, E.~Karpas,
  A.~Shashua, K.~Leyton-Brown, and Y.~Shoham, ``Parallel context windows for
  large language models,'' in \emph{Proceedings of the 61st Annual Meeting of
  the Association for Computational Linguistics (Volume 1: Long Papers)}.\hskip
  1em plus 0.5em minus 0.4em\relax Toronto, Canada: Association for
  Computational Linguistics, Jul. 2023, pp. 6383--6402.

\bibitem{ivgi2023}
M.~Ivgi, U.~Shaham, and J.~Berant, ``{Efficient Long-Text Understanding with
  Short-Text Models},'' \emph{Transactions of the Association for Computational
  Linguistics}, vol.~11, pp. 284--299, 03 2023.

\bibitem{Desislavov2023}
R.~Desislavov, F.~Martínez-Plumed, and J.~Hernández-Orallo, ``Trends in ai
  inference energy consumption: Beyond the performance-vs-parameter laws of
  deep learning,'' \emph{Sustainable Computing: Informatics and Systems},
  vol.~38, p. 100857, Apr. 2023.

\bibitem{Khattab2020}
O.~Khattab and M.~Zaharia, ``{ColBERT},'' in \emph{Proceedings of the 43rd
  International {ACM} {SIGIR} Conference on Research and Development in
  Information Retrieval}.\hskip 1em plus 0.5em minus 0.4em\relax {ACM}, Jul.
  2020.

\bibitem{Liang2022}
H.~Liang, D.~Hang, and X.~Li, ``Modeling function-level interactions for
  file-level bug localization,'' \emph{Empirical Software Engineering},
  vol.~27, no.~7, Oct. 2022.

\bibitem{Murali2021}
V.~Murali, L.~Gross, R.~Qian, and S.~Chandra, ``Industry-scale {IR}-based bug
  localization: A perspective from facebook,'' in \emph{2021 {IEEE}/{ACM} 43rd
  International Conference on Software Engineering: Software Engineering in
  Practice ({ICSE}-{SEIP})}.\hskip 1em plus 0.5em minus 0.4em\relax {IEEE}, May
  2021.

\bibitem{Razzaq2021}
A.~Razzaq, J.~Buckley, J.~V. Patten, M.~Chochlov, and A.~R. Sai,
  ``{BoostNSift}: A query boosting and code sifting technique for method level
  bug localization,'' in \emph{2021 {IEEE} 21st International Working
  Conference on Source Code Analysis and Manipulation ({SCAM})}.\hskip 1em plus
  0.5em minus 0.4em\relax {IEEE}, Sep. 2021.

\bibitem{Niu2023}
F.~Niu, C.~Mayr-Dorn, W.~K.~G. Assun\c{c}ão, L.~Huang, J.~Ge, B.~Luo, and
  A.~Egyed, ``The ablots approach for bug localization: is it replicable and
  generalizable?'' in \emph{2023 IEEE/ACM 20th International Conference on
  Mining Software Repositories (MSR)}.\hskip 1em plus 0.5em minus 0.4em\relax
  IEEE, May 2023.

\bibitem{almhana2021method}
R.~Almhana, M.~Kessentini, and W.~Mkaouer, ``Method-level bug localization
  using hybrid multi-objective search,'' \emph{Information and Software
  Technology}, vol. 131, p. 106474, 2021.

\bibitem{Sisman2012}
B.~Sisman and A.~C. Kak, ``Incorporating version histories in information
  retrieval based bug localization,'' in \emph{2012 9th {IEEE} Working
  Conference on Mining Software Repositories ({MSR})}.\hskip 1em plus 0.5em
  minus 0.4em\relax {IEEE}, Jun. 2012.

\bibitem{Shi2014}
Z.~Shi, J.~Keung, and Q.~Song, ``An empirical study of {BM}25 and {BM}25f based
  feature location techniques,'' in \emph{Proceedings of the International
  Workshop on Innovative Software Development Methodologies and
  Practices}.\hskip 1em plus 0.5em minus 0.4em\relax {ACM}, Nov. 2014.

\bibitem{Zhu2021}
Z.~Zhu, Y.~Li, Y.~Wang, Y.~Wang, and H.~Tong, ``A deep multimodal model for bug
  localization,'' \emph{Data Mining and Knowledge Discovery}, Apr. 2021.

\bibitem{Qi2022}
B.~Qi, H.~Sun, W.~Yuan, H.~Zhang, and X.~Meng, ``Dreamloc: A deep relevance
  matching-based framework for bug localization,'' \emph{IEEE Transactions on
  Reliability}, vol.~71, no.~1, p. 235–249, Mar. 2022.

\bibitem{Chakraborty2024}
P.~Chakraborty, M.~Alfadel, and M.~Nagappan, ``Rlocator: Reinforcement learning
  for bug localization,'' \emph{IEEE Transactions on Software Engineering},
  vol.~50, no.~10, p. 2695–2708, Oct. 2024.

\bibitem{Lam2017}
A.~N. Lam, A.~T. Nguyen, H.~A. Nguyen, and T.~N. Nguyen, ``Bug localization
  with combination of deep learning and information retrieval,'' in \emph{2017
  {IEEE}/{ACM} 25th International Conference on Program Comprehension
  ({ICPC})}.\hskip 1em plus 0.5em minus 0.4em\relax {IEEE}, May 2017.

\bibitem{Vancsics2022}
B.~Vancsics, F.~Horv{\'{a}}th, A.~Szatm{\'{a}}ri, and {\'{A}}.~Besz{\'{e}}des,
  ``Fault localization using function call frequencies,'' \emph{Journal of
  Systems and Software}, vol. 193, p. 111429, Nov. 2022.

\bibitem{Kim2019}
Y.~Kim, S.~Mun, S.~Yoo, and M.~Kim, ``Precise learn-to-rank fault localization
  using dynamic and static features of target programs,'' \emph{{ACM}
  Transactions on Software Engineering and Methodology}, vol.~28, no.~4, pp.
  1--34, Oct. 2019.

\bibitem{Wen2021}
M.~Wen, J.~Chen, Y.~Tian, R.~Wu, D.~Hao, S.~Han, and S.-C. Cheung, ``Historical
  spectrum based fault localization,'' \emph{{IEEE} Transactions on Software
  Engineering}, vol.~47, no.~11, pp. 2348--2368, Nov. 2021.

\bibitem{Widyasari2022}
R.~Widyasari, G.~A.~A. Prana, S.~A. Haryono, S.~Wang, and D.~Lo, ``Real world
  projects, real faults: evaluating spectrum based fault localization
  techniques on python projects,'' \emph{Empirical Software Engineering},
  vol.~27, no.~6, Aug. 2022.

\bibitem{Yang2024}
A.~Z.~H. Yang, C.~Le~Goues, R.~Martins, and V.~Hellendoorn, ``Large language
  models for test-free fault localization,'' in \emph{Proceedings of the
  IEEE/ACM 46th International Conference on Software Engineering}, ser. ICSE
  ’24.\hskip 1em plus 0.5em minus 0.4em\relax ACM, Feb. 2024.

\bibitem{Zhang2023}
Z.~Zhang, Y.~Lei, X.~Mao, M.~Yan, X.~Xia, and D.~Lo, ``Context-aware neural
  fault localization,'' \emph{IEEE Transactions on Software Engineering},
  vol.~49, no.~7, p. 3939–3954, Jul. 2023.

\bibitem{Ghanbari2023}
A.~Ghanbari, D.-G. Thomas, M.~A. Arshad, and H.~Rajan, ``Mutation-based fault
  localization of deep neural networks,'' in \emph{2023 38th IEEE/ACM
  International Conference on Automated Software Engineering (ASE)}.\hskip 1em
  plus 0.5em minus 0.4em\relax IEEE, Sep. 2023.

\bibitem{Lucia2014}
Lucia, D.~Lo, and X.~Xia, ``Fusion fault localizers,'' in \emph{Proceedings of
  the 29th ACM/IEEE International Conference on Automated Software
  Engineering}, ser. ASE ’14.\hskip 1em plus 0.5em minus 0.4em\relax ACM,
  Sep. 2014.

\bibitem{Hassan2023}
F.~Hassan, N.~Meng, and X.~Wang, ``Uniloc: Unified fault localization of
  continuous integration failures,'' \emph{ACM Transactions on Software
  Engineering and Methodology}, vol.~32, no.~6, p. 1–31, Sep. 2023.

\bibitem{Xia2014}
X.~Xia, D.~Lo, X.~Wang, C.~Zhang, and X.~Wang, ``Cross-language bug
  localization,'' in \emph{Proceedings of the 22nd International Conference on
  Program Comprehension}, ser. ICSE ’14.\hskip 1em plus 0.5em minus
  0.4em\relax ACM, Jun. 2014.

\bibitem{Li2023}
Z.~Li, H.~Zhang, X.-Y. Jing, J.~Xie, M.~Guo, and J.~Ren, ``{DSSDPP}: Data
  selection and sampling based domain programming predictor for cross-project
  defect prediction,'' \emph{{IEEE} Transactions on Software Engineering},
  vol.~49, no.~4, pp. 1941--1963, Apr. 2023.

\bibitem{yangh2024}
H.~Yang, Y.~Nong, T.~Zhang, X.~Luo, and H.~Cai, ``Learning to detect and
  localize multilingual bugs,'' 2024.

\end{thebibliography}
\balance



\end{document}